\documentclass[apj]{emulateapj}
\usepackage{amsmath}
\usepackage{color}
\usepackage[colorlinks=true,
            linkcolor=black,
            urlcolor=black,
            citecolor=blue, breaklinks=true]{hyperref}
\usepackage{url}
\usepackage{breakurl}

\shorttitle{Disk truncation in simulated cluster galaxies}
\shortauthors{Gupta et al.,}

\begin{document}

\newcommand{\tcalifa}{T-CALIFA}
\newcommand{\califa}{NT-CALIFA}
\newcommand{\rstrip}{$r_{\rm strip}$}
\newcommand{\zcl}{$\nabla Z_{\rm cl}$}
\newcommand{\zgal}{$\nabla Z_{\rm gal}$}
\newcommand{\mcl}{$\nabla M^*_{\rm cl}$}

\title{Survival of Massive Star-forming Galaxies in Cluster Cores Drives Gas-Phase Metallicity Gradients : The Effects of Ram Pressure Stripping}
\author{Anshu Gupta$^1$, Tiantian Yuan$^1$, Davide Martizzi$^2$, Kim-Vy H. Tran$^3$,  and  Lisa J. Kewley$^1$}

\affil{$^1$Research School of Astronomy and Astrophysics, Australian National University, Canberra, ACT 2611, Australia; \href{mailto:anshu.gupta@anu.edu.au}{anshu.gupta@anu.edu.au}\\  
$^2$Department of Astronomy, University of California, Berkeley, CA 94720-3411, USA\\
$^3$George P. and Cynthia W. Mitchell Institute for Fundamental Physics and Astronomy, Department of Physics \& Astronomy, Texas A\&M University, College Station, TX 77843}

\begin{abstract}

Recent observations of galaxies in a cluster at $z=0.35$ show that their integrated gas-phase metallicities increase with decreasing cluster-centric distance.  To test if ram pressure stripping (RPS) is the underlying cause, we use a semi-analytic model to quantify the ``observational bias'' that RPS introduces into the aperture-based  metallicity measurements. We take integral field spectroscopy of local galaxies, remove gas from their outer galactic disks via RPS, and then conduct mock slit observations of cluster galaxies at $z=0.35$. Our RPS model predicts a typical cluster-scale metallicity gradient of $-0.03$\,dex/Mpc.  By removing gas from the outer galactic disks, RPS introduces a mean metallicity enhancement of $+0.02$\,dex at a fixed stellar mass.  This gas removal and subsequent quenching of star formation preferentially removes low mass cluster galaxies from the observed star-forming population.  As only the more massive star-forming galaxies survive to reach the cluster core, RPS produces a cluster-scale stellar mass gradient of $-0.05\log(M_*/M_{\odot})$/Mpc. This mass segregation drives the predicted cluster-scale metallicity gradient of $-0.03$\,dex/Mpc.  However, the effects of RPS alone can not explain the higher metallicities measured in cluster galaxies at $z=0.35$.  We hypothesize that additional mechanisms including steep internal metallicity gradients and self-enrichment due to gas strangulation are needed to reproduce our observations at $z=0.35$.
\end{abstract}

%% ADD more keywords. 

\section{introduction}

Understanding the link between internal properties  of  galaxies such as stellar mass, gas fraction, chemical abundance and the large-scale environment continues to be a rich area of research. Numerous studies have established that galaxy clusters have significantly larger fractions of elliptical and S0-type galaxies relative to  the field galaxy populations  \citep{Dressler1980, Poggianti1999, VanderWel2008}. Galaxies in local galaxy clusters have lower star-formation rates and are deficient in both the molecular and atomic gas compared to field galaxies \citep{kauffmann2004, Balogh2004, Chung2009, Hughes2009, Cortese2011, Boselli2014, Odekon2016, Brown2016}. The removal of the cold/hot gas reservoir for an infalling galaxy via ram pressure stripping \citep[RPS;][]{Gunn1972}  by the hot diffuse intracluster medium (ICM) can explain some of the differences between cluster and field galaxies such as lower gas fraction and lower star formation rates in cluster galaxies \citep{Abramson2011, Chung2009, Boselli2014a}. 

The gas-phase metallicity of a galaxy  is regulated by its star-formation history and galactic gas inflows/outflows. The interactions between the ICM and circum-galactic medium (IGM/CGM) can potentially affect the chemical evolution of cluster galaxies.  Previous studies testing for environmental effects on chemical evolution have focused on measuring the stellar mass-metallicity relation (the MZ relation) variation with the local density. Observations of the MZ relation find minimal to negligible evidence of environmental influence \citep{Cooper2008, Ellison2009, Scudder2012, T.M.Hughes2012, Pasquali2012, Peng2014, darvish2015, Pilyugin2016}.  For example, \cite{Ellison2009}  and \cite{Pilyugin2016} find that galaxies in high density environments have an enhanced metallicity of $\sim 0.05$\,dex compared to field galaxies. However, both \cite{Ellison2009} and \cite{Pilyugin2016} conclude that the metal enhancement depends more strongly on the local environment density than the large-scale environment density  at redshift $z \sim 0$.

In a recent paper \cite[][hereafter Gupta16]{Gupta2016}, we measure the integrated gas-phase metallicities of cluster galaxies in MACS\,J1115+01 ($z=0.35$) and  study the correlation of the galaxy's  metallicity with the projected cluster radius.  Using this complementary method of testing for environmental effects on chemical evolution, we find that MACS\,J1115+01 has a negative cluster-scale metallicity gradient of $-0.15\,\pm\,0.08$\,dex/Mpc.  In addition, cluster galaxies are on average $\sim 0.20$\,dex more metal rich relative to field galaxies at a fixed stellar mass.  \cite{Ellison2009} also observed that galaxies at small cluster-centric distance ($< 0.3$\,R200) have a higher median metallicity than galaxies at larger cluster-centric distance ($> 0.8$\,R200) by $\sim0.03\,$dex. 

In Gupta16, we propose two scenarios that can potentially explain the observed negative cluster-scale metallicity gradient: disk truncation due to ram pressure stripping and self-enrichment due to strangulation. Self enrichment leads to the physical enhancement in the metallicity of cluster galaxies, whereas disk truncation can introduce an observational bias in the integrated metallicity measured from a fixed aperture/slit. 

Truncation of the gas disk by RPS can introduce a purely observationally-driven increase in the integrated gas-metallicities of an individual galaxy.  RPS removes the gas reservoir outside the galactic radius where the gravitational restoring force is equal to the pressure gradient exerted on the infalling galaxy by the ICM \citep{Gunn1972}.  Stripping of gas from the outer-galactic disk instantaneously suppresses the star formation in the outer part of the galaxy. The metallicity measurements via optical strong emission lines are sensitive to star-forming regions of the galaxy. Thus, the integrated emission line flux for a cluster galaxy with truncated galactic disk will be biased towards the central metal-rich part of the galaxy, resulting in an enhanced integrated metallicity measurement.  For example, \cite{T.M.Hughes2012} used disk truncation to explain the enhanced metallicity of cluster galaxies. They truncate the radial metallicity profile of NGC\,4254 from \cite{Skillman1996} to $1.5\ R_e$ and estimate a metallicity enhancement of 0.1\,dex.  However, their use of a fixed truncation radius does not consider the scatter in the truncation radius caused by different orbital parameters and physical properties of infalling galaxies.  Thus a detailed model of RPS in a cluster environment is required to better understand how disk truncation increases the observed gas-metallicities of cluster members and potentially produces cluster-scale metallicity gradients as measured by Gupta16.     

In this paper, we focus on the effect of disk truncation on the integrated metallicity of cluster members. We simulate RPS in a galaxy cluster using the semi-analytic prescription by \cite{Luo2016} and \cite{Tecce2010}. We select galaxies from  the Calar Alto Legacy Integral Field Area (CALIFA) survey data release 3 \citep{Sanchez2012} and distribute them uniformly in a galaxy cluster.  We manually truncate the 3-dimensional (3D) datacubes of  CALIFA galaxies based on the stripping radii calculated from the RPS model. We then conduct mock slit observations of the truncated CALIFA galaxies redshifted to $z=0.35$ to estimate their integrated metallicity and compare with the cluster member metallicity gradients from Gupta16.  This paper is organized as follows. In Section \ref{sec:high_z_sim}, we describe the sample selection, the mock observation technique and the metallicity estimation. Our analytic model for the ram pressure stripping is described in Section \ref{sec:disk_trunc}. In Section \ref{sec:sim}, we describe the simulation of disk truncated cluster galaxies.   In Section \ref{sec:results}, we compare our results with observations. In Section \ref{sec:discussion}, we discuss the strengths and weakness of our model. We assume a standard $\Lambda {\rm CDM\ cosmology\ with}\ \Omega_{\rm M}=0.30$, $\Omega_{\rm \Lambda}=0.70$, $\Omega_{\rm b}=0.05$, $h=0.7$.

\section{Simulating Slit Spectroscopy of Cluster Galaxies at $z\sim0.35$ }\label{sec:high_z_sim}

Disk-truncation due to RPS of outer galactic-disk introduces an observational bias in the integrated metallicity measurements via slit spectroscopy.  To determine if RPS can generate an increase in the gas-phase metallicity of cluster galaxies with decreasing cluster-centric distance, we first start with integral field unit (IFU) spectroscopy of nearby field galaxies. We can quantify the strength of the observational bias introduced in the integrated metallicity measurements by disk truncation after manually truncating the IFU datacubes using a semi-analytic model of RPS and conducting mock slit observations.

\subsection{CALIFA: IFU spectroscopy at $z\sim0$}\label{sec:sample_sel}

We select our sample from the publicly available integral field data from the third data release (DR3) of the Calar Alto Legacy Integral Field Area (CALIFA) survey \citep{Sanchez2012}. We choose the CALIFA survey because it has better spatial resolution per galaxy compared to other ongoing local IFU surveys because of the narrower redshift range \citep[$\sim 400$\,pc][]{Sanchez2016}.   Specific details of DR3 are presented in \citet{Sanchez2016} and \cite{Walcher2014}. CALIFA DR3 contains about 650 galaxies in the redshift range of $0.005<z<0.03$. The primary CALIFA sample is derived from the Sloan Digital Sky Survey  Data Release 7 \citep[SDSS DR7;][]{york2000, Abazajian2009}. We select CALIFA galaxies with reliable stellar disk scale length provided in  the \cite{J.Mendez-Abreu2016} catalog,  restricting our sample to 280 galaxies. 
Each CALIFA datacube has a spatial field of view of $74''\times 64''$ on a rectangular grid of $1''$. The average spatial resolution is $2.7''$, corresponding to an average physical resolution of $\sim1$\,kpc. The CALIFA survey has two resolution modes, low-resolution (V500) and high-resolution (V1200), with a spectral coverages of $3740-7500$\,\AA\ and $3650-4840$\,\AA\ respectively. In this study, we will use the datacubes in the low-resolution mode (V500) that have a full width half maximum (FWHM) spectral resolution of 6\,\AA\ because we need a wide spectral coverage to measure the gas-phase metallicity.

\begin{figure*}
\centering
\tiny
\includegraphics[scale=0.5, trim=0.0cm 0.0cm 0.0cm 0.0cm,clip=true]{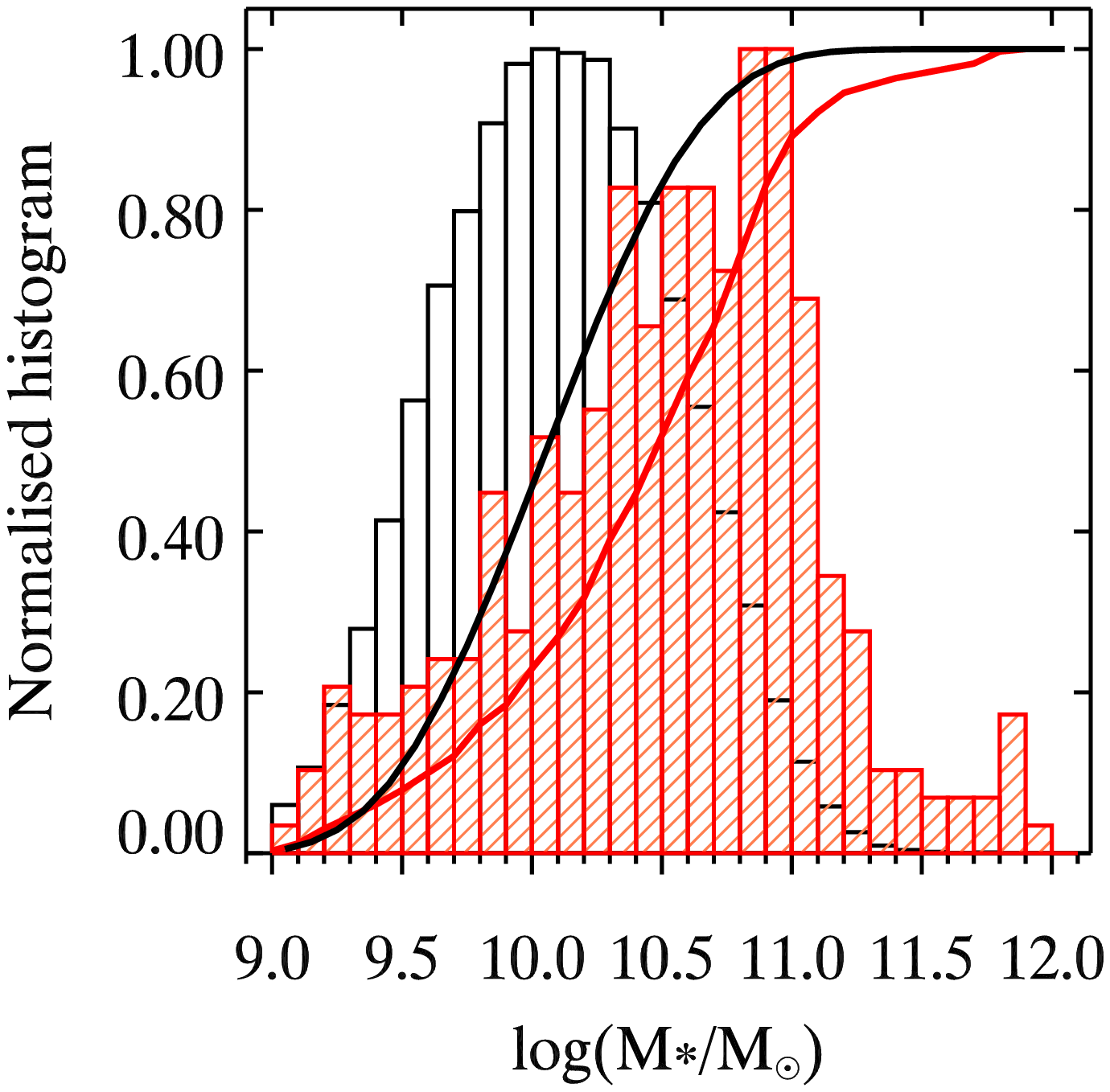}
\includegraphics[scale=0.5, trim=0.0cm 0.0cm 0.0cm 0.0cm,clip=true]{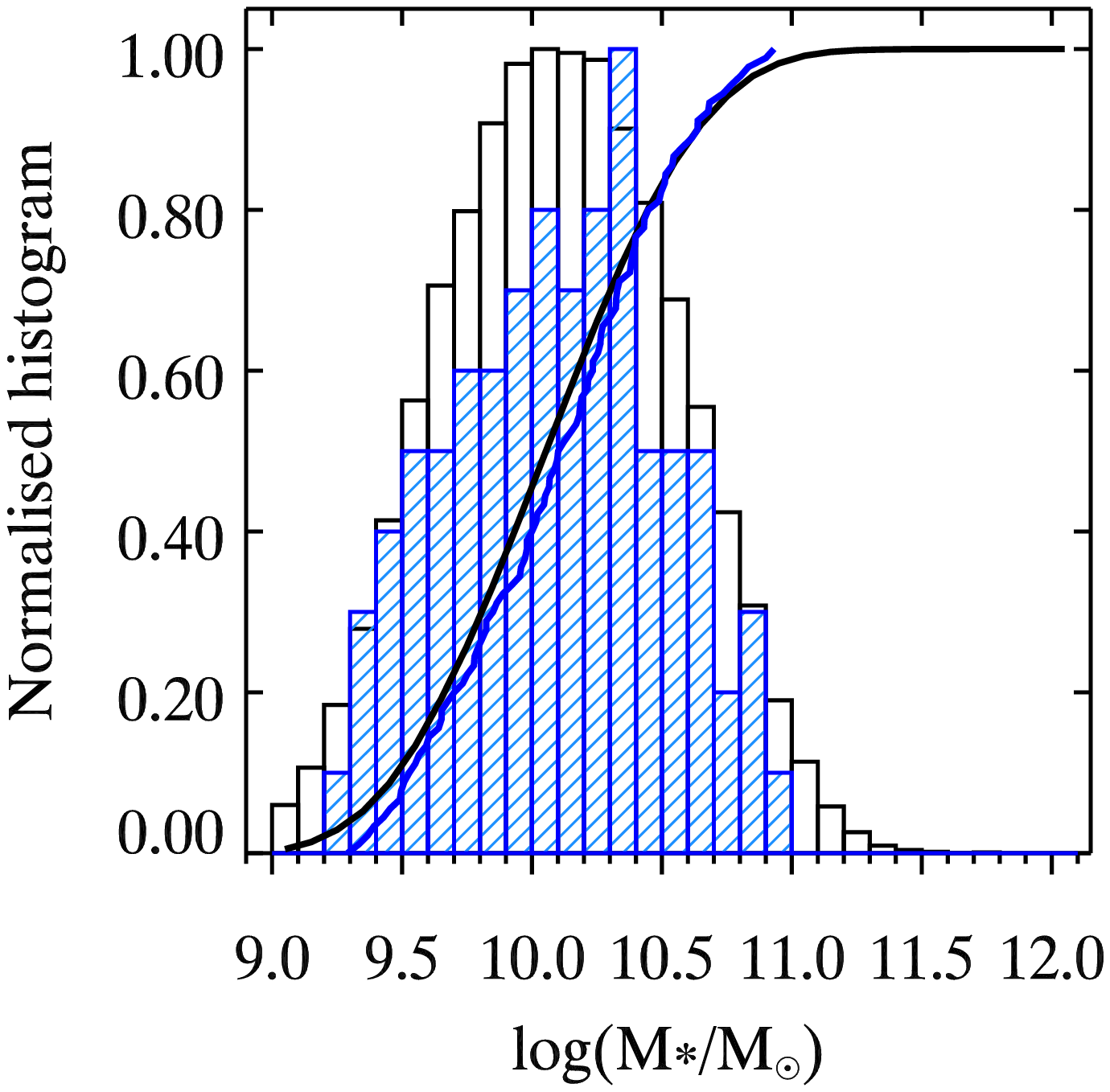}
\caption{{\bf Left panel:} The comparison of the mass distribution of the CALIFA DR3 (red histogram) with the SDSS DR7 (black histogram) sample. The CALIFA DR3 sample is biased towards the high stellar mass galaxies. The median stellar mass of full CALIFA DR3 sample is  $10^{11}  M_{\odot}$ whereas the median stellar mass of SDSS DR7 is $10^{10.5}  M_{\odot}$. {\bf Right panel:} The mass distribution of selected galaxies from the CALIFA sample (blue histogram) and SDSS DR7 (black histogram). We select CALIFA galaxies by using the stellar mass function of SDSS DR7 and populating it with CALIFA galaxies. The stellar mass distribution of  our selected CALIFA galaxies and SDSS DR7 have a KS-test probability of 0.99 to be drawn from the same parent sample.   }
\label{fig:mass_dis}
\end{figure*}

To simulate disk truncation in cluster galaxies, we require a representative sample of field galaxies. The SDSS DR7 survey has $\sim 1$ million galaxies and typically represents field galaxies. Using the stellar mass estimates by \cite{GonzalezDelgado2014} for the CALIFA sample, we find that the CALIFA DR3 sample is biased towards higher stellar mass relative to the full SDSS DR7 data (Figure \ref{fig:mass_dis}: left panel). Stellar masses of the SDSS DR7 galaxies are taken from the MPA-JHU catalogs \citep{Kauffmann2003a, Brinchmann2004}. We generate the stellar mass function of SDSS DR7 galaxies and populate it with CALIFA DR3 galaxies. Due to the limited number of galaxies with stellar masses below $10^{9.1}  M_{\odot}$ in the CALIFA sample, we only match the stellar mass function at stellar mass  $M_*> 10^{9.1}  M_{\odot}$.  The blue histogram in the right panel of Figure \ref{fig:mass_dis} shows the stellar mass distribution of the selected CALIFA galaxies. A two sided Kolmogorov--Smirnov (KS) test yield a probability of 99\% for the stellar mass distribution of the SDSS sample and the selected CALIFA sample to be derived from the same parent population. Our final sample consists of $91$ galaxies in the mass range of ${\rm 9.29<\log(M_*/M_{\odot})<10.93}$. 

\subsection{Redshifting CALIFA to $z_{\rm cl}=0.35$}\label{sec:redshifting}

The clusters observed in Gupta16 were  at a redshift of $z_{\rm cl} \sim 0.35$, whereas the average redshift of the selected CALIFA sample is $z\sim0.013$. To redshift the CALIFA galaxies to $z_{\rm cl} \sim0.35$, we use  basic cosmology equations of angular size variation and surface brightness dimming.   The mean redshift of the selected CALIFA sample is $z\sim0.013$, leading to an average reduction in angular size and total flux by a factor of $0.054$ and $9\times 10^{-4}$ respectively.  

\begin{figure*}
\centering
\tiny
\includegraphics[scale=0.7, trim=0.0cm 4.0cm 0.0cm 0.0cm,clip=true]{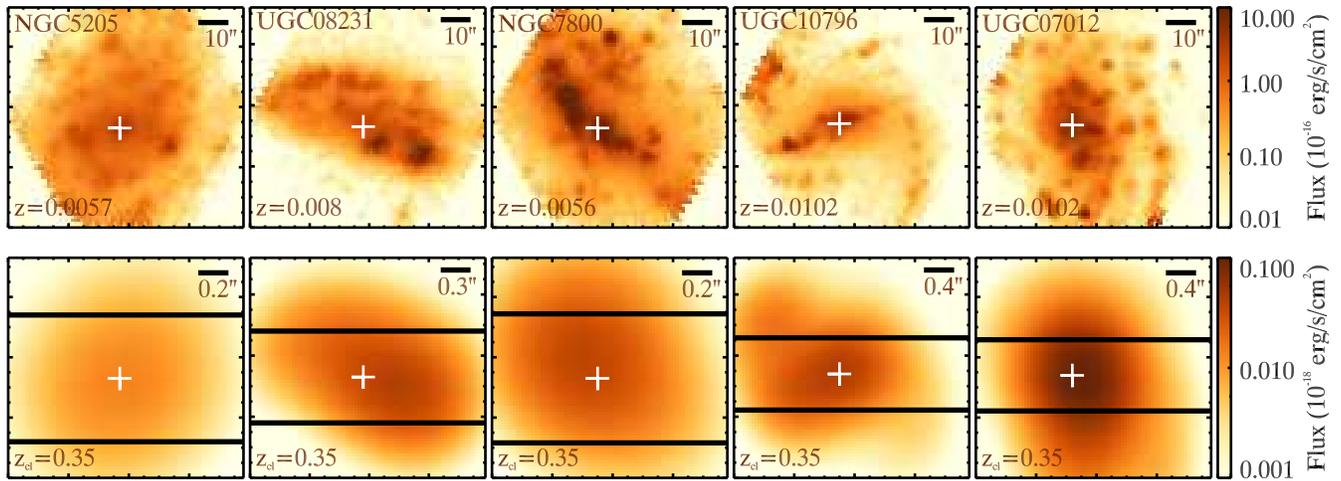}
\caption{{\bf Top panel:} Wavelength-collapsed H$\alpha$ intensity map for a random sample of 5 galaxies from the stellar mass-matched CALIFA sample at $z\sim0$. {\bf Bottom panel:} Redshifted, wavelength-collapsed H$\alpha$ intensity map at $z_{\rm cl} = 0.35$, with a seeing of $0.''8$.  The thick black lines in each panel represent the $5''\times 1''$ DEIMOS slit. The angular scale before and after redshifting is marked on the top right of each panel. To generate a 1D slit spectrum, we integrate the flux inside the slit for each wavelength slice in the datacube. }
\label{fig:ha_map}
\end{figure*}

The observations in Gupta16 were conducted with the DEIMOS spectrograph on the Keck-II telescope, which is a multi-slit spectrograph in seeing-limited conditions. To consider the effect of beam smearing in high-redshift observations, we match the spatial resolution of the redshifted datacubes with  DEIMOS/Keck  observations. The seeing of DEIMOS observations in Gupta16 was $0.''8$, corresponding to a spatial resolution of 3.952\,kpc at $z_{\rm cl}$. The average seeing of CALIFA data is $2.''7$, with a spatial sampling of $1''$. We convolve each wavelength slice in redshifted datacubes with the point spread function (PSF) of DEIMOS observations. The top panel in Figure \ref{fig:ha_map} shows the wavelength-collapsed H$\alpha$  intensity maps  for five randomly selected galaxies from the CALIFA sample and the corresponding H$\alpha$  intensity map in the bottom panel after redshifting the datacubes to $z_{\rm cl} = 0.35$ and convolving with a seeing of $0.''8$ .

\subsection{Mock slit spectroscopy}\label{sec:slit_obs}

We estimate the metallicity of cluster galaxies in Gupta16 using integrated spectra from Keck/DEIMOS observations. To make a realistic comparison with the cluster-scale abundance gradient observed in Gupta16, we conduct mock DEIMOS observations of the selected CALIFA galaxies after redshifting them to $z_{\rm cl}$.  We generate a single slit spectrum for each redshifted IFU datacube by positioning a $5''\times 1''$ DEIMOS slit on the galaxy center (Figure\,\ref{fig:ha_map}: bottom panel) and integrating the flux within the slit.

\subsection{Emission line fitting and metallicity estimation}\label{sec:met_est}

We extract emission line fluxes  by following the same procedure as described in Gupta16. In summary, we fit multiple Gaussian profiles to the 1D slit spectrum. The initial redshift  is taken from the CALIFA catalog. We extract emission line fluxes of the following strong emission lines: H$\beta$, [O\,III]\,$\lambda\,4959,\,5007$, H$\alpha$, [N\,II]\,$\lambda\,6583$ and [S\,II]\,$ \lambda\lambda\,6717,6731$. We remove galaxies with signal-to-noise (S/N) $< 3$ for any emission line. The AGN galaxies are removed  using the [O\,III]/H$\beta$ and [N\,II]/H$\alpha$ diagram and  the theoretical calibration of \cite{Kewley2001}.  

The metallicity of the star-forming galaxies is derived using the new diagnostic by \cite{Dopita2016} based on the ratio of [N\,II]/H$\alpha$ and [N\,II]/[S\,II]. The metallicity calibration by \cite{Dopita2016} is given by: 
\begin{equation}
\begin{split}
\mathrm{12 + \log(O/H) = 8.77 + \log([N\,II]/[S\,II])} + 
\\ {\rm 0.264\log([N\,II]/H\alpha)}.
\end{split}
\end{equation}
Out of the 91 galaxies in the mass-selected CALIFA sample, we can reliably measure the metallicity of 61 galaxies after implementing the S/N cut and removing AGN galaxies.  

\section{modeling the effect of disk truncation on gas metallicity measurements}\label{sec:disk_trunc}

RPS leads to many physical changes in properties of galaxies in the cluster environment. The atomic and molecular gas are observed trailing behind infalling cluster galaxies, giving direct evidence of RPS in cluster environments \citep{Abramson2011, Boselli2014a, DeGrandi2016}. Observations of jellyfish galaxies show extreme stripping events where even the dense star-forming regions are stripped \citep{Owen2006, Cortese2007, Owers2012, McPartland2015, Monroe2017}. The gas removal by RPS can explain why cluster galaxies are HI deficient compared to field galaxies and why the HI deficiency increases as the galaxy moves closer to the cluster center \citep{Hughes2009}. The HI disk for field galaxies extends beyond the stellar disk \citep{Cormier2016}, whereas the HI disk in cluster galaxies is usually smaller or the same size as the stellar disk \citep{Chung2009, Boselli2014a}. The gas removal due to RPS in cluster environment has been used to explain the formation of  S0-type galaxies in rich galaxy clusters \citep{Sun2006, Aragon-Salamanca2006, Bedregal2011}. The RPS leads to reduction in total mass and size of the  gaseous disk of satellite galaxies, a phenomenon we  refer to as disk truncation. 

Disk galaxies in the local universe have a universal negative internal metallicity gradient \citep{Zaritsky1994, Sanchez2013, Ho2015}. If the progenitor of a cluster satellite exhibited a negative internal metallicity gradient before falling into the cluster, its integrated metallicity after disk truncation might be modified because of the removal of the low-metallicity outer portions of the disk. This picture is somewhat simplified because it ignores satellite-satellite interactions and residual star formation in the satellites that may lead to the production of more metals. However, in our treatment, we will neglect satellite-satellite interactions and residual star formation and focus on the zeroth order effect of disk truncation.
Our goal is to test  whether a simple disk-truncation model is sufficient to explain the metallicity gradient for cluster satellites observed in Gupta16. 

\subsection{Our disk truncation model}

Studying disk truncation in cluster satellites in its full complexity is very challenging because the simulation of full physics of multi-phase interstellar medium (ISM) and its hydrodynamical interaction with the ICM requires sophisticated modeling and can only be achieved using expensive, high-resolution numerical simulations \citep{Roediger2005, Steinhauser2016, Luo2016}. Hydrodynamical simulations use RPS of cold gas to obtain the star formation rate suppression and color transformation in simulated galaxy clusters \citep{Abadi1999, Balogh2000, Roediger2005, Roediger2007, HeB2012, Steinhauser2016}.  

Semi-analytic models of galaxy formation usually have a simplistic description of the RPS of the gaseous component in cluster galaxies. Most semi-analytic models follow the prescription by \cite{Gunn1972}  for both the hot and the cold gas to calculate the stripping radius and the mass loss \citep{McCarthy2008, Tecce2010, Guo2011, Luo2016}. Comparisons  with hydrodynamical simulations suggest that most semi-analytic models overestimate the strength of RPS, producing a larger fraction of red galaxies in clusters \citep{Guo2013, wang2014, Steinhauser2016}.  To test the effect of disk truncation on integrated metallicity observations, we choose a simplified approach that allows us to explicitly parameterize the physics of disk truncation.

We use an analytic model of disk truncation that depends only on the orbital parameters of the satellite galaxy and the local ICM properties. The procedure is  summarized as follows:
\begin{enumerate}
\item We assume that the mass-selected CALIFA galaxies (for details see \S\ \ref{sec:sample_sel}) are uniformly distributed in the cluster as a first order initial condition for RPS to act upon (further discussed in Section \ref{sec:assumptions}). We assign the orbital velocity to each galaxy using the radial velocity distribution for MACS\,J1115+01 from Gupta16 (\S\ \ref{sec:sim}). 

\item We use the analytic model to calculate the stripping radius for each galaxy $r_{\rm strip}$ (\S\ \ref{sec:rp_mod}). We then  ``simulate" disk truncation for all galaxies, i.e., we create mock datacubes for galaxies where the flux at radii $r>r_{\rm strip}$ is set to zero (\S\ \ref{sec:sim} ).

\item We conduct mock DEIMOS observations of the simulated disks and measure the integrated metallicity for each simulated cluster galaxy  (\S\ \ref{sec:met_est}). We calculate the cluster-scale metallicity gradient following the same procedure of Gupta16.

\item  The above procedure generates one realisation  of the disk truncation simulation with a cluster-scale metallicity gradient measurement. We repeat the above process until the mean cluster scale metallicity gradient converges (Figure \ref{fig:met_grad_dis}), which  requires a minimum of 150 realisations.
\end{enumerate}

Our procedure does not take into account the full dynamics of the disk interaction with the ICM. We will discuss the ramifications of our assumptions in Section 6.  Our current approach, albeit simple, provides a useful framework to understand the physical effect of disk truncation on the cluster galaxy metallicity. 

\subsection{An analytic model for the ram pressure stripping of cold gas}\label{sec:rp_mod}

The multi-phase ISM of a galaxy can be divided into a cold and a hot component, primarily composed of the molecular hydrogen and neutral/ionized hydrogen, respectively. Explicit modeling of the stripping of different phases of the ISM requires properly designed numerical studies and is beyond the scope of our model.  Both numerical simulations and observations show that the warmer neutral/ionized gas is stripped first \citep{McCarthy2008, Chung2009, Boselli2014a, Steinhauser2016}, but the stripping of the hot gas would not instantaneously affect the star formation rate because of large cooling timescales. The stripping of the cold gas component from the outer disk of a galaxy would quench the star formation almost immediately, resulting in the loss of emission line flux from the stripped part of the galactic disk. Thus, the stripping of the cold gas from the outer parts of the galactic disk  will directly affect the  average metallicity measurement for the satellite galaxy. 

We follow the prescription by \cite{Gunn1972} to model the RPS of gas in a galaxy falling into the ICM.  We assume that the cold gas is fully stripped beyond the galacto-centric radius at which the internal gravitational potential energy density of the satellite galaxy is balanced by the RP.  We define $r_{\rm strip}$ as the galactic radius at which the ISM pressure is equal to the RP exerted by the ICM. The RP experienced by a galaxy at a cluster-centric distance of $R$ falling into the ICM  with an orbital velocity of $v_{orb}$ is 
\begin{equation}\label{eq:rp}
P_{r.p.} = \rho_{\rm ICM} (R) v_{orb}^2,
\end{equation} 
where $\rho_{\rm ICM} (R)$ is the ICM density at $R$. The cold gas is gravitationally bound with the ISM of the galaxy. Thus, the cold gas will be stripped only if the RP exceeds the gravitational restoring force due to the ISM ($P_{\rm ISM}$). We use Equation \eqref{eq:pism} given by \cite{Tecce2010} to determine the $P_{\rm ISM}$ 
\begin{equation}\label{eq:pism}
P_{\rm ISM} = 2\pi G[\Sigma_*(r) + \Sigma_g(r)] \Sigma_g(r),
\end{equation}
where $\Sigma_*(r)$ is the stellar surface mass density and $\Sigma_g(r)$ is the surface mass density of the gas.
At $r_{\rm strip}, P_{r.p} = P_{\rm ISM}$, and thus we can equate \eqref{eq:rp} and \eqref{eq:pism},  for a galaxy at a cluster-centric distance of $R$
\begin{equation}\label{eq:rp_strip} 
2\pi G[\Sigma_*(r_{\rm strip}) + \Sigma_g(r_{\rm strip})] \Sigma_g(r_{\rm strip}) =  \rho_{\rm ICM} (R) v_{orb}^2.
\end{equation}
To estimate $r_{\rm strip}$ we need to model the stellar and gas surface mass density profile ($\Sigma_*(r)$ \& $\Sigma_g(r)$), and the ICM density profile ($\rho_{\rm ICM} (R) $).  The cold gas disk at radius $>r_{\rm strip}$ from the galactic center will be stripped away. 

We assume that both the stellar disk and the gas disk follow exponential surface density profiles
\begin{equation}\label{eq:star}
\Sigma_*(r) = \Sigma_{*0} \exp(-r/r_*)  
\end{equation}
and 
\begin{equation}\label{eq:gas}
\Sigma_g(r) = \Sigma_{g0} \exp(-r/r_g),   
\end{equation}
where $r_*$ and $r_g$ are the scale length of stellar and gas disks, respectively, and $\Sigma_{*0}$ and $\Sigma_{g0}$ are the corresponding central surface mass density. The  total stellar mass of an exponential surface density disk is given by $M_* = 2\pi r_*^2\Sigma_{*0}$ and the total gas mass is $M_g = 2\pi r_g^2\Sigma_{g0}$. We assume the same scale length for both the stellar and the cold gas disk, i.e., $r_* = r_g$, which is corroborated by the spatially resolved observations of the molecular gas for local field galaxies \citep{Bigiel2012, Cormier2016}. The ratio of central surface mass density of the gas and the stellar disk is equal to the molecular gas fraction ($M_g/M_* $) of the galaxy, 
\begin{equation}\label{eq:fgas}
\frac{\Sigma_{g0}}{\Sigma_{*0}} =  \frac{M_g}{M_*} = f_{gas}.
\end{equation}
The molecular gas fraction for local field galaxies depends weakly on the stellar mass and has a typical value of $f_{gas} = 0.1$  \citep{Boselli2014}. Thus, for a complete model of the gas and stellar surface density profiles, we only require the total stellar mass $M_*$ (obtained from the CALIFA catalogs) and the scale length of stellar disk $r_*$. We use  \cite{J.Mendez-Abreu2016} catalog to obtain $r_*$ for CALIFA DR3 galaxies. 

We evaluate the RP stripping radius ($r_{\rm strip}$)  in Equation  \eqref{eq:rp_strip}, using the modelled stellar and gas surface density profile (Equation \eqref{eq:star} to \eqref{eq:fgas}) as
\begin{equation}\label{eq:r_strip}
r_{strip} = -\frac{r_*}{2}\ln\left[\frac{\rho_{\rm ICM} (R) v_{orb}^2}{2\pi G \Sigma_{*0}^2f_{gas}(1+f_{gas})}\right].
\end{equation} 
The modelled stripping radius (\rstrip) depends weakly on the ICM density ($\rho_{\rm ICM}$) and the orbital velocity ($v_{orb}^2$) of the satellite galaxy. A similar model of RPS has been used in various semi-analytic models of galaxy evolution \citep{DeLucia2004,Tecce2010,Guo2011}.  

Different models can be used to generate the ICM density profile ($\rho_{\rm ICM}$) of the cluster. We use the \cite{Makino1998} model to generate the ICM density profile of MACS J1115+01. According to \cite{Makino1998}, if the dark matter halo of the cluster follows the Navarro, Frenk and White \citep[NFW; ][]{Navarro1997} density profile,
\begin{equation}
\rho_{DM} (R) =  \frac{\rho_s}{(R/R_s)(1+R/R_s)^2},
\end{equation}  
where $\rho_s$ and $R_s$ are the scale density and scale length; then the ICM density is given by: 
\begin{equation} \label{eq:icm_den}
\rho_{\rm ICM} (R)  = \rho_{g0} e^{-B} \left(1+\frac{R}{R_s}\right)^{(BR_s/R)},
\end{equation}
where $\rho_{g0}$ is the central gas density and B is the ratio of the gravitational energy to the kinetic energy. In this formalism, B is given by:
\begin{equation}
B = \frac{4\pi G \mu m_p \rho_s R_s^2}{kT_X},
\end{equation}
where $\mu$ is the mean particle weight, $m_p $ is the proton mass and $T_X$ is the temperature of the hot ICM gas. The central gas density ($\rho_{g0}$) is given by, 
\begin{equation}\label{eq:rho_g0}
\begin{split}
\rho_{g0} = \frac{f_{g} \Omega_b\rho_s}{\Omega_m} e^B\left[\ln(1+c) - \frac{c}{1+c}\right] \\
 \times \left[\int_0^c x^2(1+x)^{B/x} dx\right]^{-1},
\end{split}
\end{equation}
again taken from \cite{Makino1998}, where $f_g$ is the ratio of the gas mass to the total baryonic fraction in the cluster and $c$ is the concentration parameter given by the ratio of the virial radius to the scale length.  Observationally, $f_g$ ranges from $0.7-0.9$ \citep{Gonzalez2013}. We adopt $f_g = 0.8$.  Thus, we require only four parameters, $\rho_s$ (scale density of NFW profile), $R_s$ (scale length of NFW profile), $kT_X$ (X-ray temperature) and $c$ (concentration parameter) to determine the density profile of the hot ICM. To compare with the cluster-scale metallicity gradient observations for MACS\,J1115+01,  we use the cluster parameters for MACS\,J1115+01 provided by \cite{Merten2014} to model the density profile of ICM ($\rho_s = 0.61\times 10^{15}\ h^2 M_{\odot} {\rm Mpc}^{-3}$, $R_s = 0.62\ {\rm Mpc}/h$, $kT_X = 8.0$ keV and $c = 2.9$). Thus, the stripping radius in Equation \eqref{eq:r_strip} only depends on  $R$ and $v_{orb}$. Before constructing a more realistic model of disk truncation in a galaxy cluster, we investigate the effect of the internal metallicity gradient on the cluster-scale metallicity gradient using an idealised approach in next subsection (\S\ \ref{sec:fiducial}). 

\begin{figure}
\centering
\tiny
\includegraphics[scale=0.5, trim=0.0cm 0.0cm 0.0cm 0.0cm,clip=true]{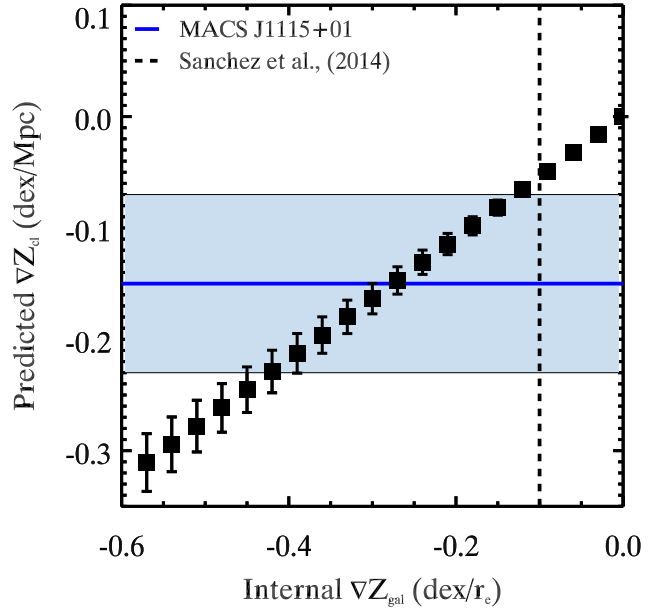}
\caption{ The predicted cluster-scale metallicity gradient (\zcl) as a function of the  internal metallicity gradient (\zgal) of the infalling galaxy from the idealised model of disk truncation described in Section \ref{sec:fiducial}.  The blue solid line and shaded region represent the the mean and $1\sigma$ error for the \zcl\  observed for MACS\,J1115+01 in Gupta16. The black dashed line represent the characteristic \zgal\  for the CALIFA sample from \cite{Sanchez2013}. For infalling galaxies with a constant \zgal\,$= -0.1\,{\rm dex/r_e}$, our RPS model predict a \zcl\,$ = -0.05\,$dex/Mpc. }
\label{fig:fid_met_grad}
\end{figure}

\begin{figure*}
\centering
\tiny
\includegraphics[scale=0.7, trim=0.0cm 0.0cm 0.0cm 0.0cm,clip=true]{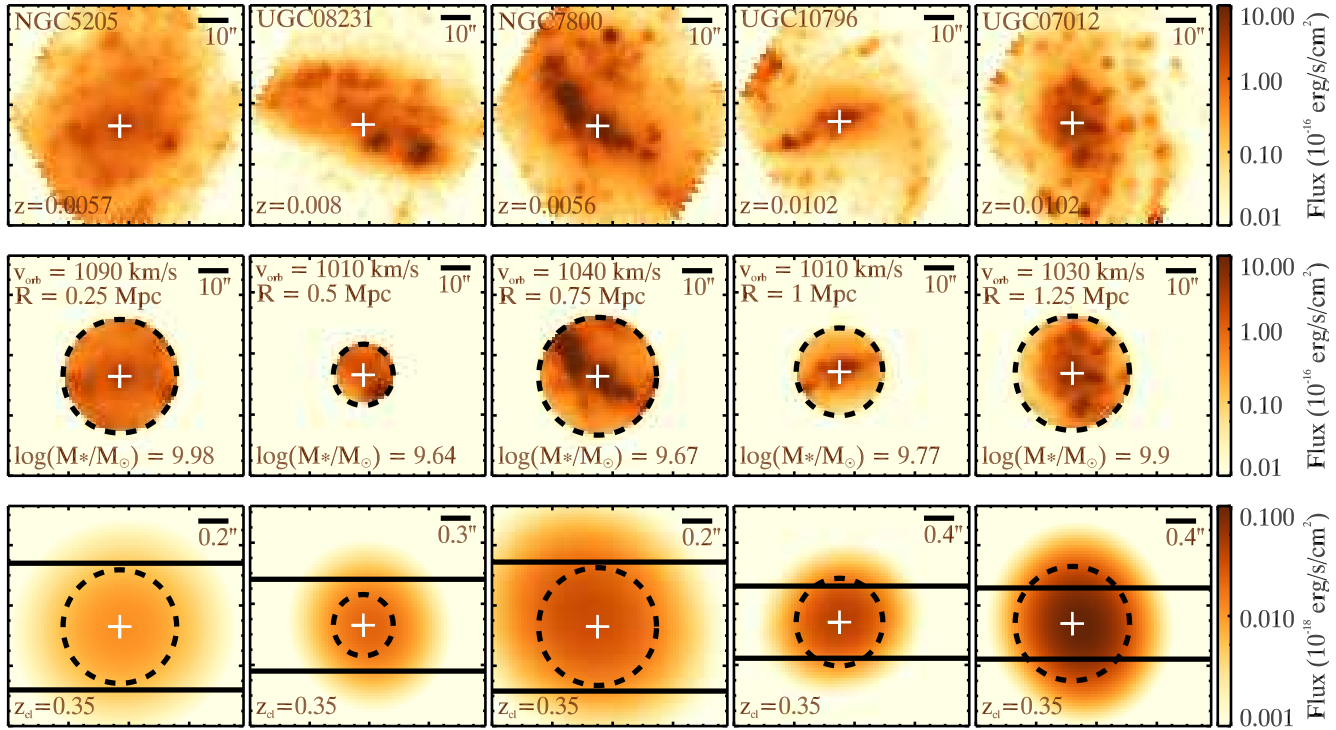}
\caption{{\bf Top panel:} The wavelength-collapsed H$\alpha$ intensity map for 5 CALIFA galaxies (Same as top panel in Figure \ref{fig:ha_map}). {\bf Middle panel:} Truncated H$\alpha$ intensity map, where the stripping radius is calculated using $R$ and $v_{orb}$ specified in each panel. The dashed-circle represents the truncation region. {\bf Bottom panel:} The truncated H$\alpha$ intensity map  after redshifting galaxies to $z_{\rm cl} = 0.35$, with a seeing of $0.''8$.  The black lines represents the $5''\times 1''$ DEIMOS slit. To generate a 1D-slit spectra, we integrate the flux inside the DEIMOS slit.}
\label{fig:ha_map_trunc}
\end{figure*}

\subsection{Internal metallicity gradients vs. cluster-scale gradients}\label{sec:fiducial}

IFU observations of local field galaxies indicate the existence of a universal negative internal metallicity gradient for local disk galaxies \citep{Zaritsky1994, Sanchez2013, Ho2015}. The observation of metallicity enhancement via  truncation of outer-galactic disk is based on the assumption that all galaxies have a negative internal metallicity gradient.  To quantify the effect of an internal metallicity gradient on cluster-scale metallicity gradient, we manually place a galaxy with fixed internal properties (such as stellar mass and internal metallicity gradient) at different cluster-centric distances.  

We define \zcl\ as the cluster-scale gradient in the integrated metallicity  of star-forming cluster galaxies and \zgal\ as the internal metallicity gradient of a satellite galaxy.   We assume that a $10^{10}  M_{\odot}$ galaxy with a particular \zgal\  is falling into the cluster with an orbital velocity of $1000\,$km/s. The empirical stellar mass-size relation is used to estimate the stellar disk scale length for this $10^{10}  M_{\odot}$ galaxy \citep{Mosleh2013}. We manually place the satellite galaxy  at different cluster-centric distance and estimate the stripping radius at each cluster-centric distance using our RPS model (Equation\,\eqref{eq:r_strip}). We estimate the integrated metallicity ($Z_{\rm int}(R)$) at each cluster-centric distance by taking an intensity-weighted average of \zgal\ within \rstrip\ using following equation:
\begin{equation}\label{eq:met_int}
Z_{\rm int}(R) = \int_0^{r_{\rm strip}(R)} \exp(-r/r_e) \nabla Z_{\rm gal} dr,
\end{equation}
where we assume an exponentially declining profile for the intensity \citep{T.M.Hughes2012}.  We perform a linear fit between $Z_{\rm int}(R)$ and the cluster-centric distance to  estimate \zcl.  To estimate the relation between \zcl\ and \zgal, we assume that the internal metallicity gradient of the infalling galaxy varies between  $-0.6 \le \nabla Z_{\rm gal} \le 0.0\,{\rm dex/r_e}$ \citep{Yuan2011, Jones2013a, Ho2015} and measure \zcl\ for each value of the \zgal.

Figure \ref{fig:fid_met_grad} shows that \zcl\ is directly proportional to the \zgal\ of the satellite galaxy, i.e., $\nabla Z_{\rm cl} = 0.54 \times \nabla Z_{\rm gal}$. Our RPS model predicts a \zcl\ of $-0.05$\,dex/Mpc, when adopting a $\nabla Z_{\rm gal} = 0.1\,{\rm dex/}r_e$,  characteristic for CALIFA galaxies  \citep{Sanchez2013}. By using the CALIFA sample to simulate cluster galaxies at  $z_{\rm cl} = 0.35$, we neglect the possible evolution in the internal metallicity gradients of star-forming galaxies out to a redshift of $z_{\rm cl} = 0.35$. 

The evolution of the internal metallicity gradient of star-forming galaxies with redshift is contentious. Star-forming galaxies at $z\sim0$ have a universal negative internal metallicity gradient \citep{Sanchez2013, Ho2015}. The adaptive-optics observations of lensed galaxies show that the internal metallicity gradient steepens with redshift \citep{Yuan2011, Jones2013a}. In contrast, seeing limited observations of non-lensed galaxies find a flat metallicity gradient for  high redshift galaxies \citep{Wuyts2016}. \cite{Yuan2013} show that low spatial resolution in seeing limited conditions leads to gradient flattening in the high-redshift observations. If \zcl\ observed for MACS\,J1115+01 in Gupta16 is caused only by the truncation of outer-galactic disks, then we predict an  \zgal\,$= -0.28\,{\rm dex/r_e}$ for MACS\,J1115+01 cluster members in this idealised model. This internal gradient is significantly steeper than any observations of high-redshift galaxies and is unlikely to be real. Nevertheless, the variation in \zgal\ would introduce scatter in the observed \zcl\ and  would be interesting to conduct observations to measure \zgal\ for MACS\,J1115+01 cluster members.

Using a single galaxy with fixed internal properties disregards the scatter introduced in the predicted \zcl\ due to the scatter in internal properties of  infalling galaxies. To realistically simulate disk truncation in MACS\,J1115+01, we use a stellar  mass selected sample of CALIFA galaxies.

\begin{figure*}
\centering
\tiny
\includegraphics[scale=0.65, trim=0.0cm 0.0cm 0.0cm 0.0cm,clip=true]{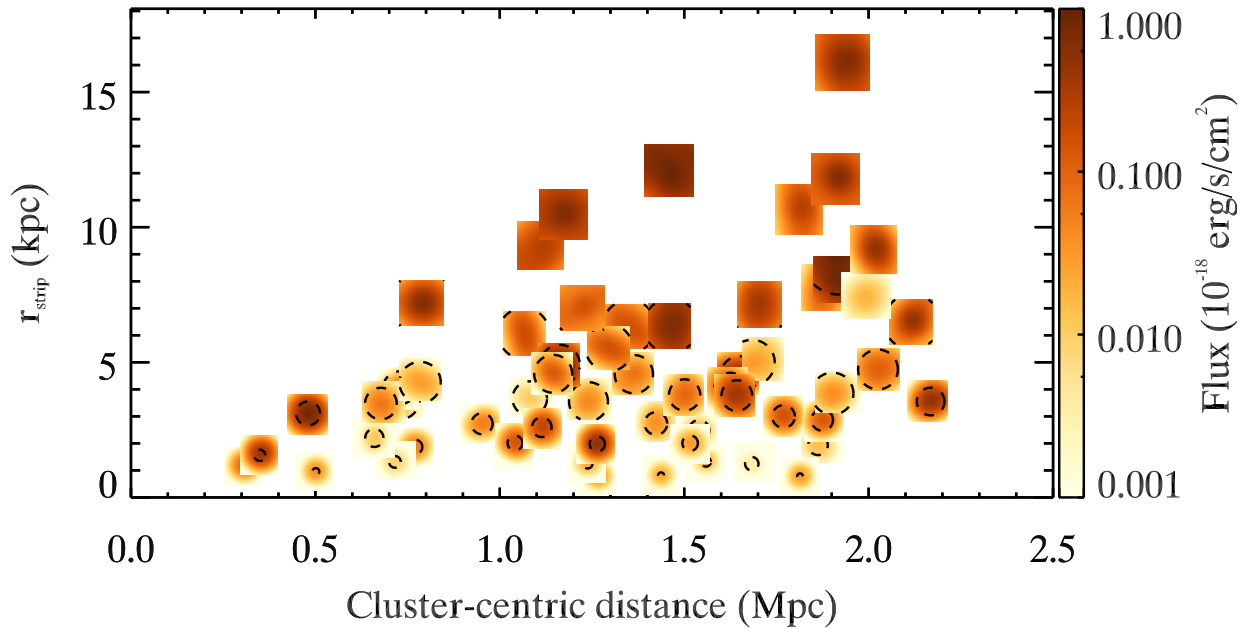}
\caption{The variation of stripping radius ($r_{\rm strip}$) versus cluster-centric distance for the selected CALIFA galaxies. The stripping radius is calculated by assuming a uniform distribution of galaxies in the cluster and the Gaussian distribution of the orbital velocity. Galaxies with $r_{\rm strip}$ greater than 0.5 kpc are shown on the figure. Each data points corresponds to the wavelength-collapsed H$\alpha$ map of the galaxies at $z_{\rm cl} = 0.35$ and observed with a seeing of $0.8''$. The dashed circle in each H$\alpha$ map defines the stripping boundary corresponding to the $r_{\rm strip}$. The \rstrip\ shows a clear positive correlation with the cluster-centric distance, i.e., the size of galactic disk (where ISM emission line/cold gas can be detected) decreases near the cluster center. The large scatter in the stellar mass of infalling cluster galaxies results in a large scatter in the \rstrip\ at a particular cluster-centric distance.}
\label{fig:disk_size}
\end{figure*}

\section{Simulating disk truncation in a galaxy cluster}\label{sec:sim}

We simulate disk truncation in a galaxy cluster by uniformly distributing the mass selected CALIFA galaxies at different cluster-centric radii. The stripping radius for each galaxy is evaluated using Equation \eqref{eq:r_strip}. To compare with the observations from Gupta16, we model the ICM density profile using Equations \eqref{eq:icm_den} to \eqref{eq:rho_g0}, where the cluster parameters were obtained from \cite{Merten2014} for MACS\,J1115+01. We neglect the central 200\,kpc region in the cluster because all galaxies closer than 200\,kpc will be completely stripped of the cold gas according to Equation \eqref{eq:r_strip}. Note that our results do  not change significantly  by using a minimum cluster-centric radius of  500\,kpc as used in Gupta16. 

The observations in Gupta16 give us the velocity dispersion along the line-of-sight ($\sigma_{\rm los} =960\,$km/s) for the cluster members.  To assign orbital velocity, we assume that our simulated galaxy cluster is isotropic, i.e., all three components of velocities have the same Gaussian distribution with dispersion equal to $\sigma_{\rm los}$.  The orbital velocity is given by $v_{\rm orb} = \sqrt{v_x^2+v_y^2+v_z^2}$. Thus, $v_{\rm orb}$ follow a Maxwell-Boltzmann distribution with mean = 1532\,km/s  and dispersion = 646\,km/s. We assign the orbital velocity independent of a galaxy's position in the cluster.

\subsection{Galaxies with truncated galactic disks}\label{sec:t-califa}

We compare the \zcl\ observation from Gupta16 to \zcl\  observed for the mass matched sample from CALIFA DR3. We define \tcalifa\ as the sample of cluster galaxies at $z_{\rm cl} = 0.35$ with truncated galactic disks for which metallicity is estimated via the mock DEIMOS observations. The stripping radius \rstrip\  for each  \tcalifa\ galaxy is estimated  using Equation \eqref{eq:r_strip} after assigning orbital parameters, i.e., cluster-centric radius and orbital velocity.  To simulate the disk truncation, we set the flux at $r>r_{\rm strip}$  to zero for each wavelength slice, thereby suppressing the emission line flux outside the stripping radius. We emphasize that our disk truncation model removes only the cold gas at $r > r_{\rm strip}$. Other components of the galaxy such as stars, dust and the halo are intact  (\S\ \ref{sec:mass_grad}). However, because we are simulating observed emission line fluxes, it is reasonable to cut off data at $r > r_{\rm strip}$ because no emission line would be detected at $r > r_{\rm strip}$. After disk truncation, we redshift the galaxies to $z_{\rm cl}$ (\S\ \ref{sec:redshifting}) and conduct mock observations with  DEIMOS (\S\ \ref{sec:slit_obs}). The metallicity is estimated using emission line ratios [N\,II]/H$\alpha$ and [N\,II]/[S\,II] (\S\ \ref{sec:met_est}) extracted from the 1D-slit spectrum for each \tcalifa\ galaxy. 

Figure \ref{fig:ha_map_trunc} gives a figurative description of our disk truncation simulation for a random sample of 5 galaxies. The truncation radius for each galaxy is calculated after assigning the galaxy an orbital velocity and distance from the cluster center. The middle row in Figure \ref{fig:ha_map_trunc} shows the truncated galactic disks after reassigning the flux outside $r_{\rm strip}$ to zero. After truncating the galactic disk, we redshift the galaxies to $z_{\rm cl} =0.35$ (\S\ \ref{sec:redshifting}) and convolve with the PSF of $0.''8$ (Figure \ref{fig:ha_map_trunc}: bottom panel). We truncate the galactic disk before redshifting to avoid spatial resolution effects in disk truncation. The 1D spectrum of the truncated galaxy is extracted by integrating the flux within a  $5''\times 1''$ DEIMOS slit (\S\ \ref{sec:slit_obs}) at each wavelength slice. Figure \ref{fig:disk_size} shows the relation between \rstrip\ and cluster-centric distance for the full mass  selected CALIFA sample. The mean stripping radius gets smaller as cluster galaxies move closer to the cluster center.  The stellar mass of infalling galaxies varies almost 2 orders in magnitude, resulting in a large scatter in the \rstrip\ of galaxies at a particular cluster-centric distance.

\begin{figure*}
\centering
\tiny
\includegraphics[scale=0.33, trim=0.5cm 0.5cm 0.0cm 2.0cm,clip=true]{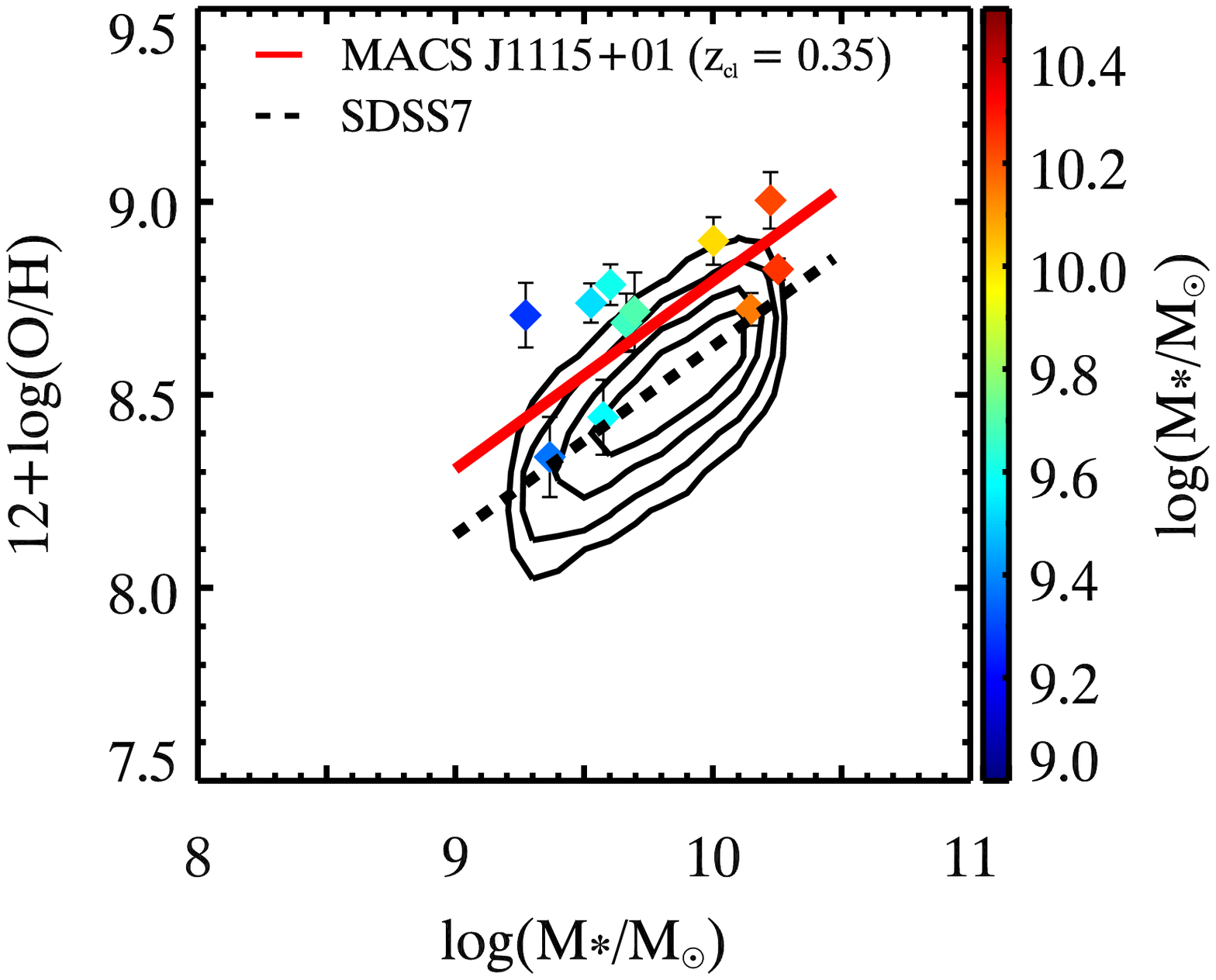}
\includegraphics[scale=0.33, trim=0.5cm 0.5cm 0.0cm 2.0cm,clip=true]{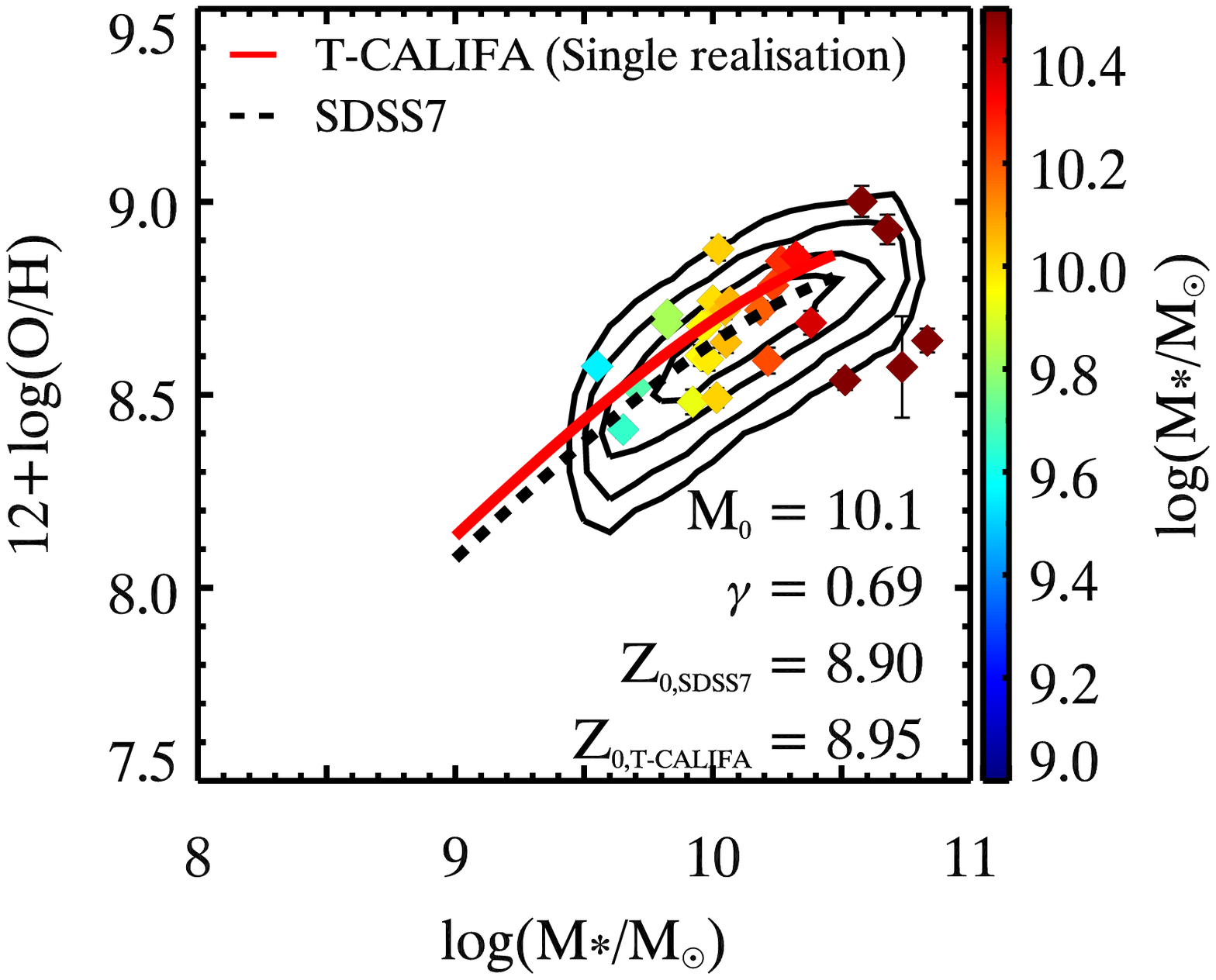}
\includegraphics[scale=0.33, trim=0.5cm 0.5cm 0.0cm 2.0cm,clip=true]{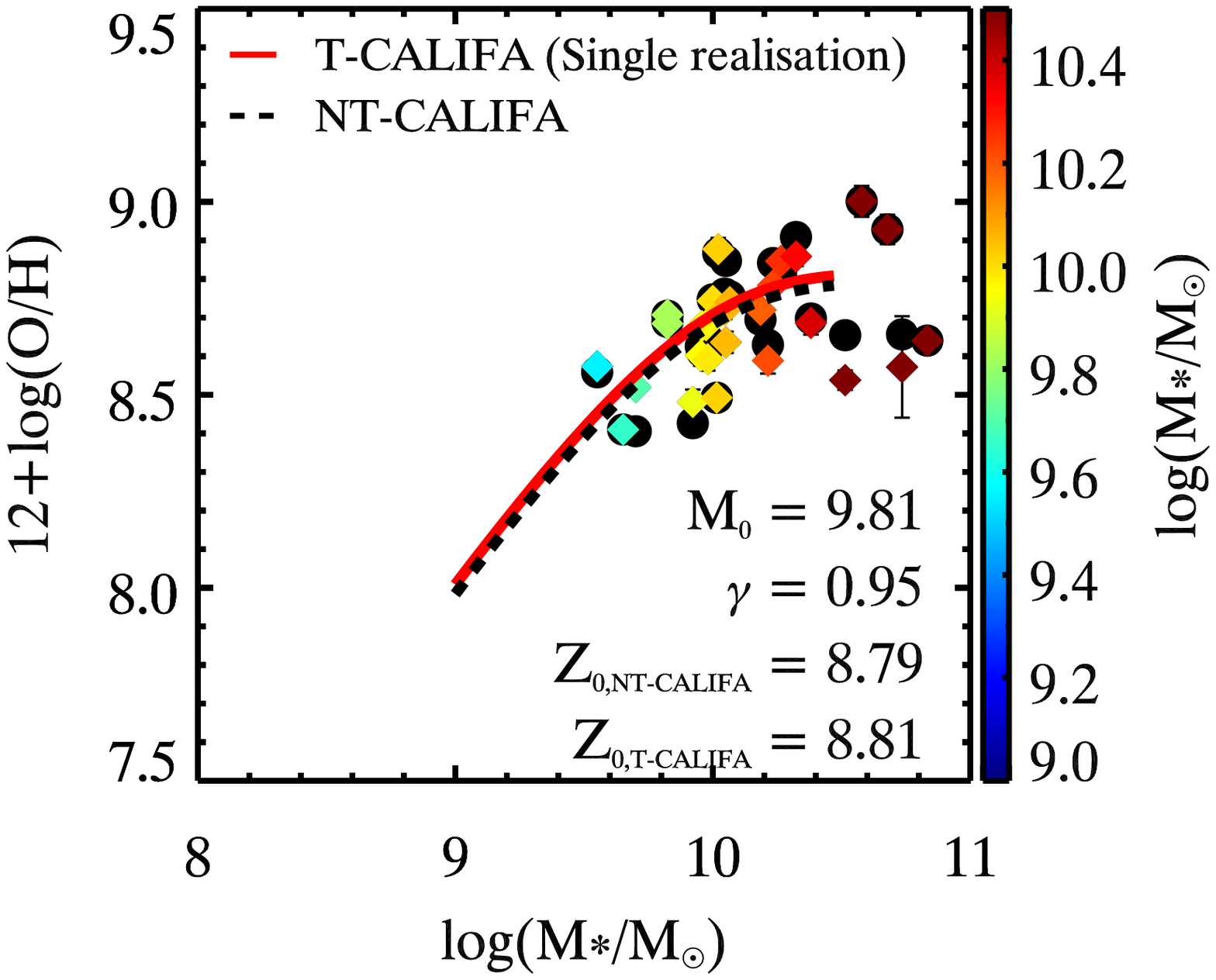}
\caption{ {\bf Left panel:} The MZ relation comparison between  MACS\,J1115+01 and the local SDSS galaxies (Figure 6 in Gupta16).  We observe an metallicity enhancement of $+0.20\,\pm\,0.13$\,dex/Mpc for cluster members at a fixed stellar mass. {\bf Middle panel:} The comparison of MZ relation for \tcalifa\ galaxies with the SDSS7 sample for one realisation of disk truncation simulation. The contours in left and middle panel  represents the MZ relation for the local SDSS galaxies (25\%, 50\%, 70\% and 90\%). {\bf Right panel:} The MZ relation comparison between \tcalifa (diamonds) and \califa (black circles) galaxies sample for one realisation of disk truncation simulation.  The cluster data points in each panel are color coded with stellar mass. The red solid and black dashed  line represents the best linear fit to MZ relation of  the cluster galaxies sample and the respective comparison sample in each panel. The fit parameters are indicated in the figure legend, where $Z_{\rm 0,T-CALIFA}$ is the best-fit asymptotic metallicity for \tcalifa\ galaxies, $Z_{\rm 0,SDSS7}$ is the best-fit asymptotic metallicity for the SDSS7 sample and   $Z_{\rm 0,NT-CALIFA}$ is the best-fit asymptotic metallicity for the \califa\ sample.   The difference between the best-fit asymptotic metallicity to the \tcalifa\ sample and the control sample represents the metallicity offset due to disk truncation. }
\label{fig:mz}
\end{figure*}

\begin{figure*}
\centering
\tiny
\includegraphics[scale=0.5, trim=0.5cm 0.5cm 0.0cm 2.0cm,clip=true]{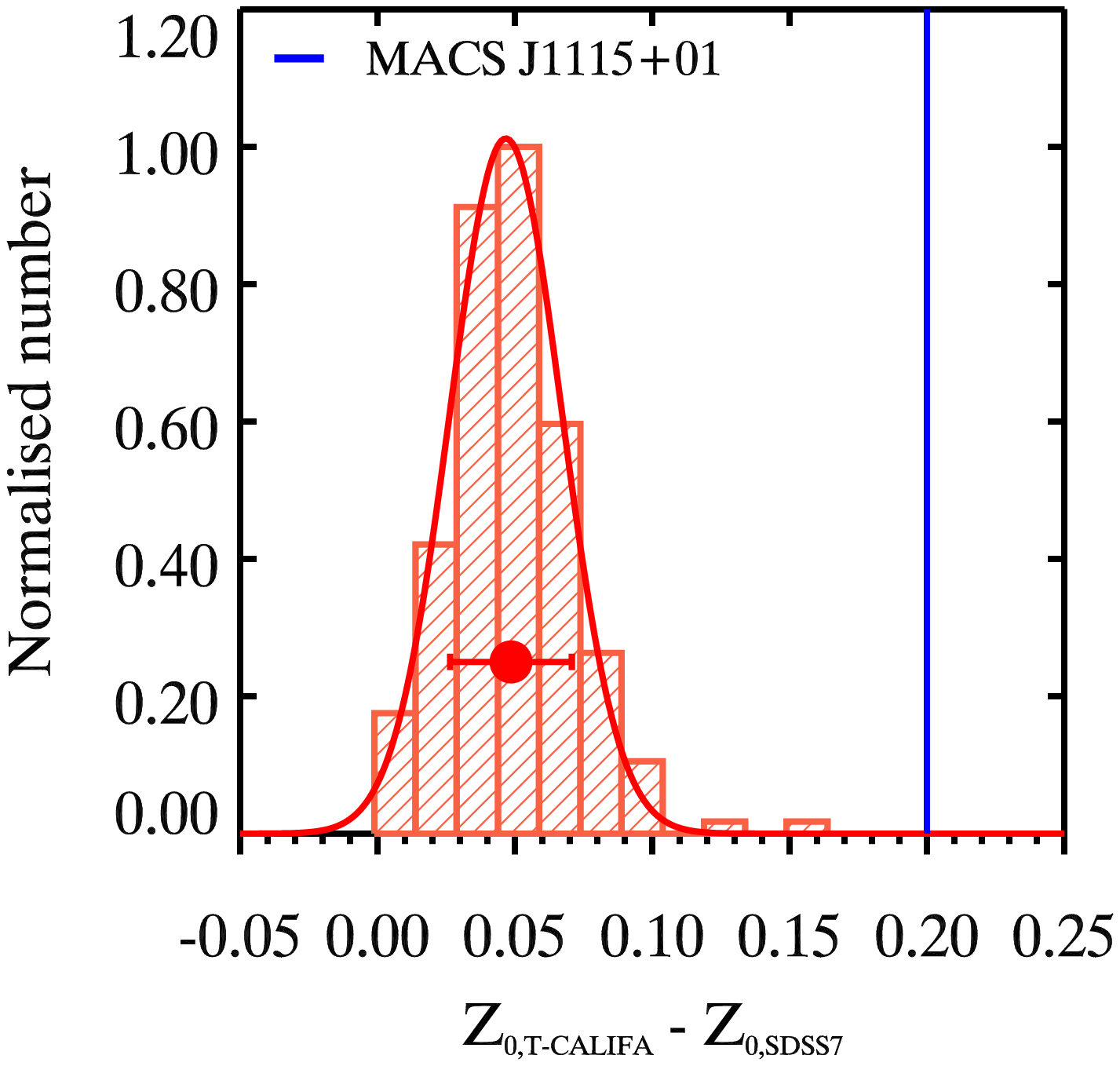}
\includegraphics[scale=0.5, trim=0.5cm 0.5cm 0.0cm 2.0cm,clip=true]{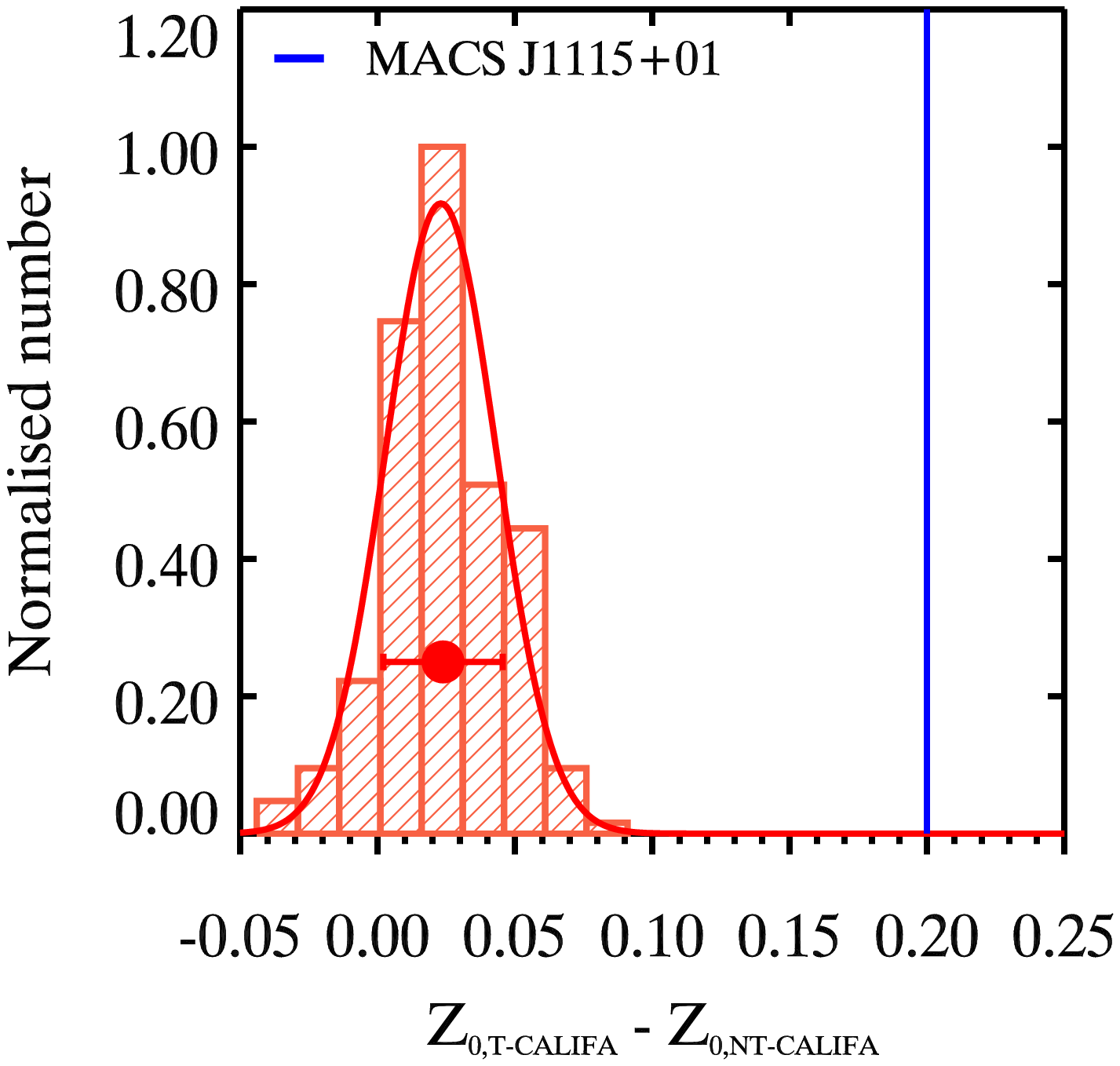}
\caption{
The distribution of metallicity offset between the MZ relation for the \tcalifa\ galaxies and the control samples using the all realisations of the disk truncation simulation. {\bf Left panel}: Comparison with SDSS7 sample. The red circle represents the mean ($+0.05$\,dex) and the 1-$\sigma$ scatter (0.02\,dex)  in the metallicity offset distribution.  {\bf Right panel} Comparison with \califa\ sample.  The red circle represents the mean (+0.02\,dex) and the 1-$\sigma$ scatter (0.02\,dex)  in the metallicity offset distribution. The solid blue line in each panel represents the metallicity enhancement at a fixed stellar mass observed for MACS\,J1115+01 in Gupta16 (Figure \ref{fig:mz}: left panel). Disk truncation leads to a mean metallicity enhancement of $0.02$\,dex, but the metallicity enhancement from disk truncation is significantly less than the metallicity enhancement observed for MACS\,J1115+01 ($0.20\,\pm\,0.13$\, dex).}
\label{fig:mz_hist}
\end{figure*}

\subsection{Comparison sample}\label{sec:comp_sample}

To estimate the metallicity enhancement due to disk truncation, we compare the mass-metallicity relation for the \tcalifa\ galaxies with the following control samples:

\subsubsection{SDSS7 (Fiber spectroscopy)}\label{sec:sdss7}
To have a direct comparison with the metallicity offset in the mass-metallicity relation observed in Gupta16, we use the SDSS DR7 galaxy sample as one of our control sample. The stellar mass and emission line flux measurements for SDSS DR7 galaxies are taken from the publicly available MPA-JHU catalogs \footnote{ \label{fn:1} See \url{http://wwwmpa.mpa-garching.mpg.de/SDSS/DR7/Data/stellarmass.html}} \citep{Kauffmann2003a, Brinchmann2004}. We use the same selection criteria  as Gupta16 to ensure a consistent comparison sample. In brief,  we remove galaxies with aperture coverage of $<20\%$ leading to a redshift selection of $0.05<z<0.1$ and remove AGN galaxies using the Baldwin-Phillips-Terlevich (BPT) diagram \citep{Baldwin1981, Phillips1986, Veilleux1987} procedure  described in \cite{Kewley2006}.  We further remove galaxies with signal-to-noise (S/N) ratio $< 3.0$ for any of the following emission lines:  [O\,II]\,$\lambda\lambda\,3726,\,3729$, H$\beta\,\lambda\,4861$, [O\,III]\,$\lambda\lambda\,4959,\,5007$, [N\,II]\,$\lambda\lambda\,6549,\,6583$, H$\alpha\,\lambda\,6563$ and [S\,II]\,$\lambda\lambda\,6717,\,6731$. 

\subsubsection{Non-truncated CALIFA sample}\label{sec:nt-califa}

We define \califa\ as the mass selected CALIFA galaxy sample redshifted to $z_{\rm cl}$ for which no disk truncation is performed. We extract the mass-metallicity relation for \califa\ sample to directly compare the effect of disk truncation on metallicity. We use the same stellar mass matched sample of galaxies from the full CALIFA DR3 catalog as used for the \tcalifa\, sample,  redshift them to $z_{\rm cl} =0.35$, extract 1D slit spectra and measure the global metallicity again using  emission line ratios [N\,II]/H$\alpha$ and [N\,II]/[S\,II].  The \califa\ sample undergo the exact sample treatment as the \tcalifa\ sample except for the disk truncation. The metallicity enhancement observed between \tcalifa\ and \califa\ samples can only be attributed to the bias introduced in the metallicity measurement because of disk truncation. 

\section{Comparison with observations}\label{sec:results}

We extract the integrated metallicity of disk truncated cluster galaxies at $z_{\rm cl}$ (\tcalifa; \S\ \ref{sec:t-califa}).  To measure the metallicity we use emission line ratios [N\,II]/[S\,II] and [N\,II]/H\,$\alpha$ and  apply a S/N cut of 3 for all necessary emission lines. Additionally, we remove galaxies that have an AGN or the star-formation rate is less than $0.2\ M_{\odot}/{\rm yr}$ (\S\ \ref{sec:met_est}).  \cite{J.Mendez-Abreu2016} provide measurement of $r_*$  for nearly 280 CALIFA DR3 galaxies  out of the full CALIFA DR3 sample of 650 galaxies. The stripping radii estimation requires accurate measurement of $r_*$, further limiting the sample size of galaxies with reliable gas-phase metallicity estimates.  To estimate the scatter introduced by the RP model, we perform multiple realisations of the disk truncation simulation, each time redistributing the mass matched CALIFA galaxies in the galaxy cluster and reassigning the orbital velocity based on the Gaussian random distribution. The mean cluster-scale metallicity gradient converges in about 150 realisations of disk truncation simulation, but to avoid underestimation of scatter we repeat our disk truncation simulation for 200 realisations.   Out of the 91 galaxies in the mass matched CALIFA sample, we are left with 25-30 galaxies per realisation for which metallicities could be determined.

\begin{figure*}
\centering
\tiny
\includegraphics[scale=0.5,  trim=0.5cm 0.8cm 0.0cm 1.4cm,clip=true]{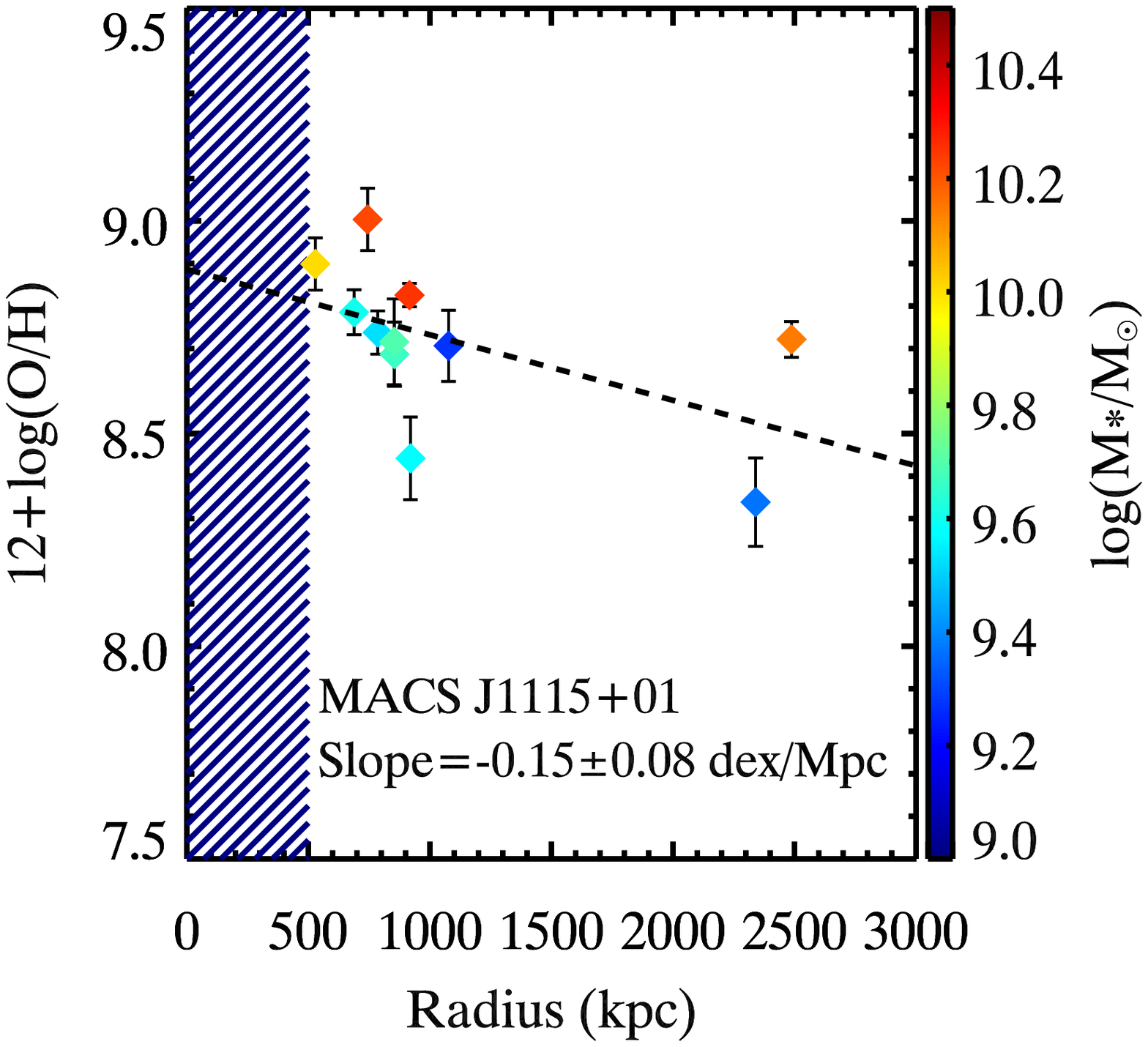}
\includegraphics[scale=0.5,  trim=0.5cm 0.8cm 0.0cm 1.4cm,clip=true]{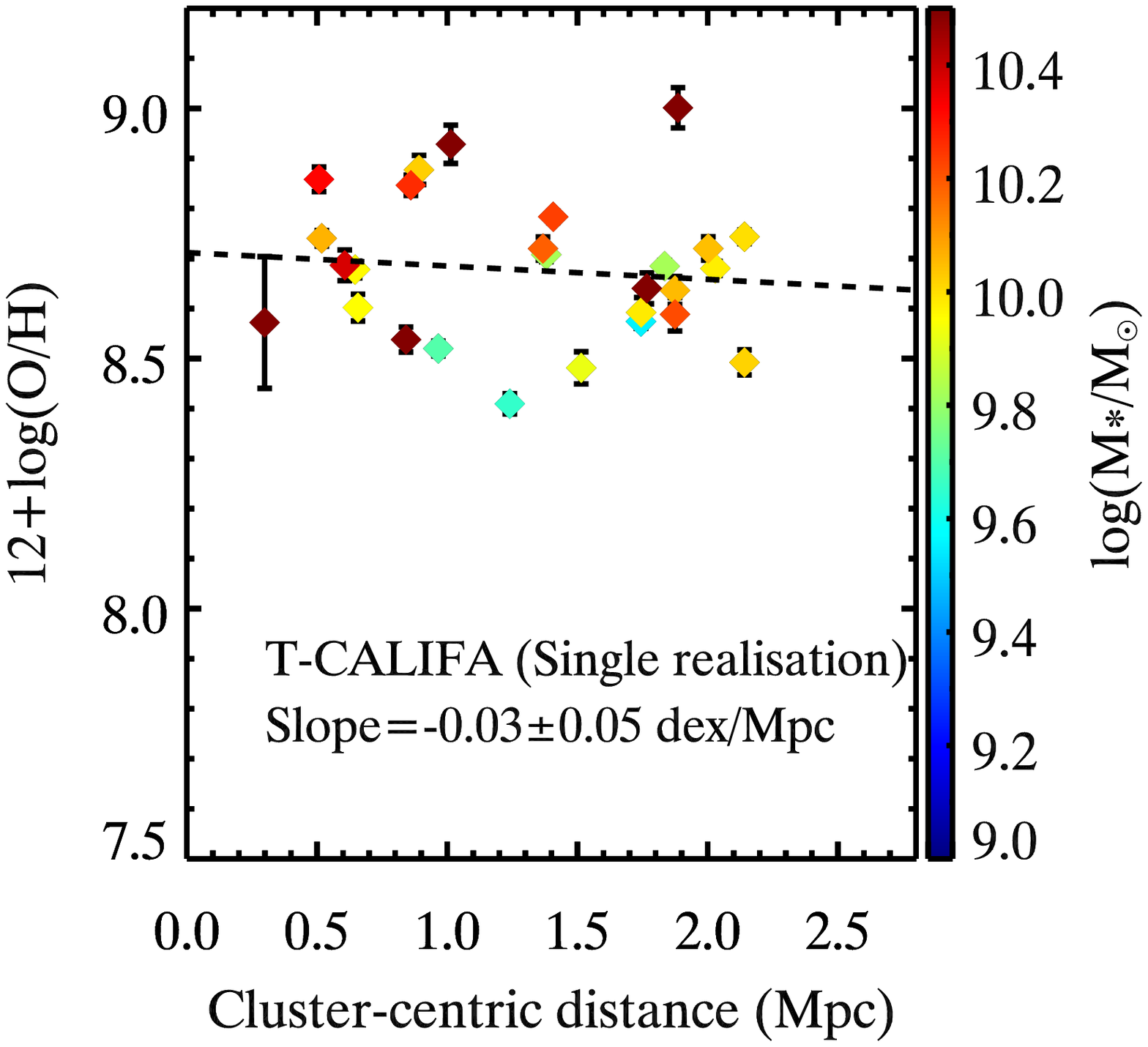}
\caption{{\bf Left panel:} The distribution of gas-phase metallicity of cluster members in MACS\,J1115+01 with respect to projected distance from the cluster center (Figure 7 in Gupta16).  {\bf Right panel:} The distribution of integrated metallicity of star-forming galaxies in \tcalifa\ sample with the cluster-centric distance for one realisation of disk truncation simulation. Each data point in both panels  is color coded with the stellar mass. The dashed line in each panel represents the best linear fit to the data and the best-fit slope is labeled in the panel. We extract the intercept metallicity and the \zcl\  for each realisation of disk truncation simulations to understand the distribution of \zcl\ predicted by our disk truncation simulation. }
\label{fig:met_dis}
\end{figure*}

\begin{figure*}
\centering
\tiny
\includegraphics[scale=0.5,  trim=0.5cm 0.5cm 0.0cm 2.0cm,clip=true]{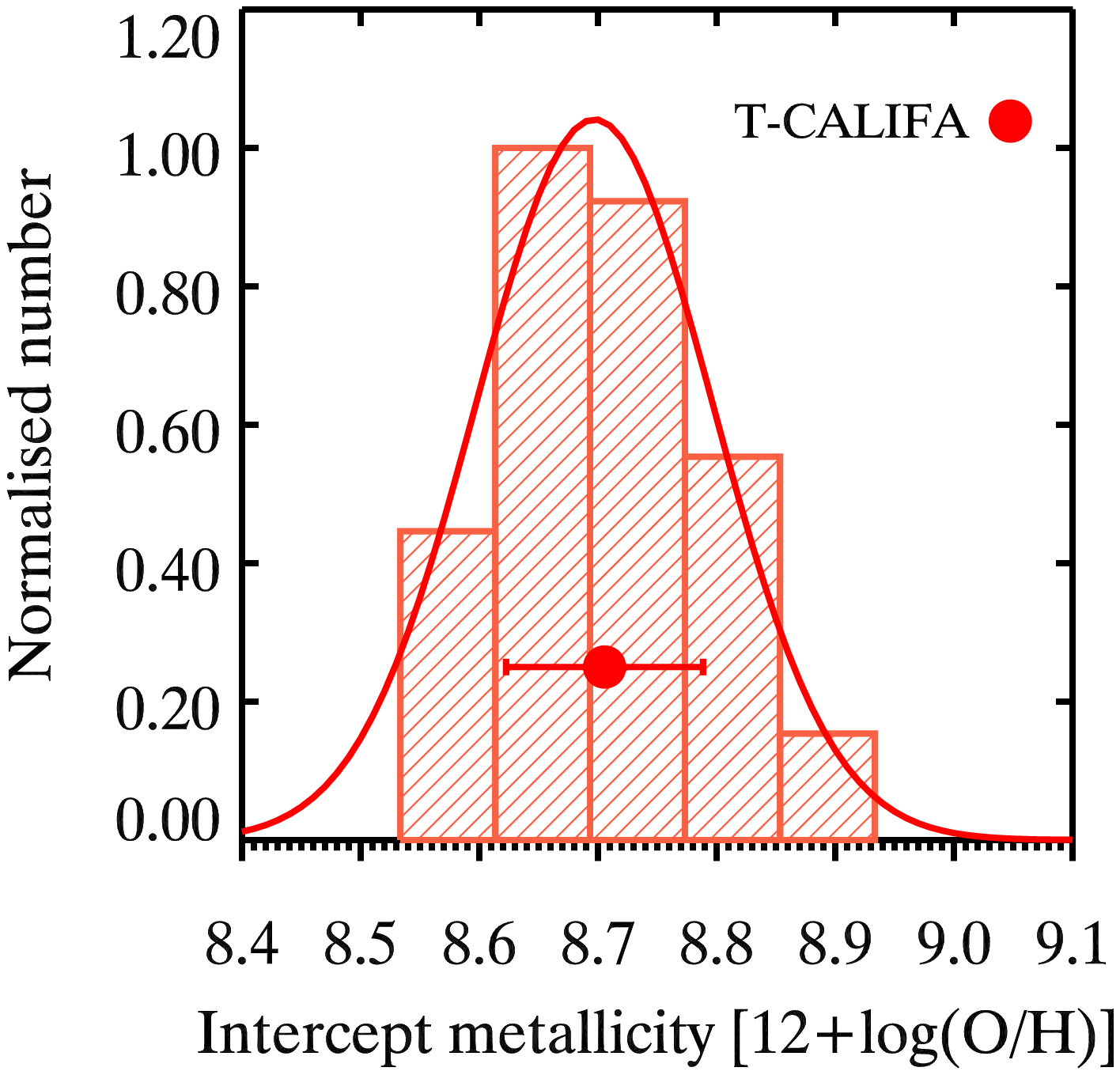}
\includegraphics[scale=0.5,  trim=0.5cm 0.5cm 0.0cm 2.0cm,clip=true]{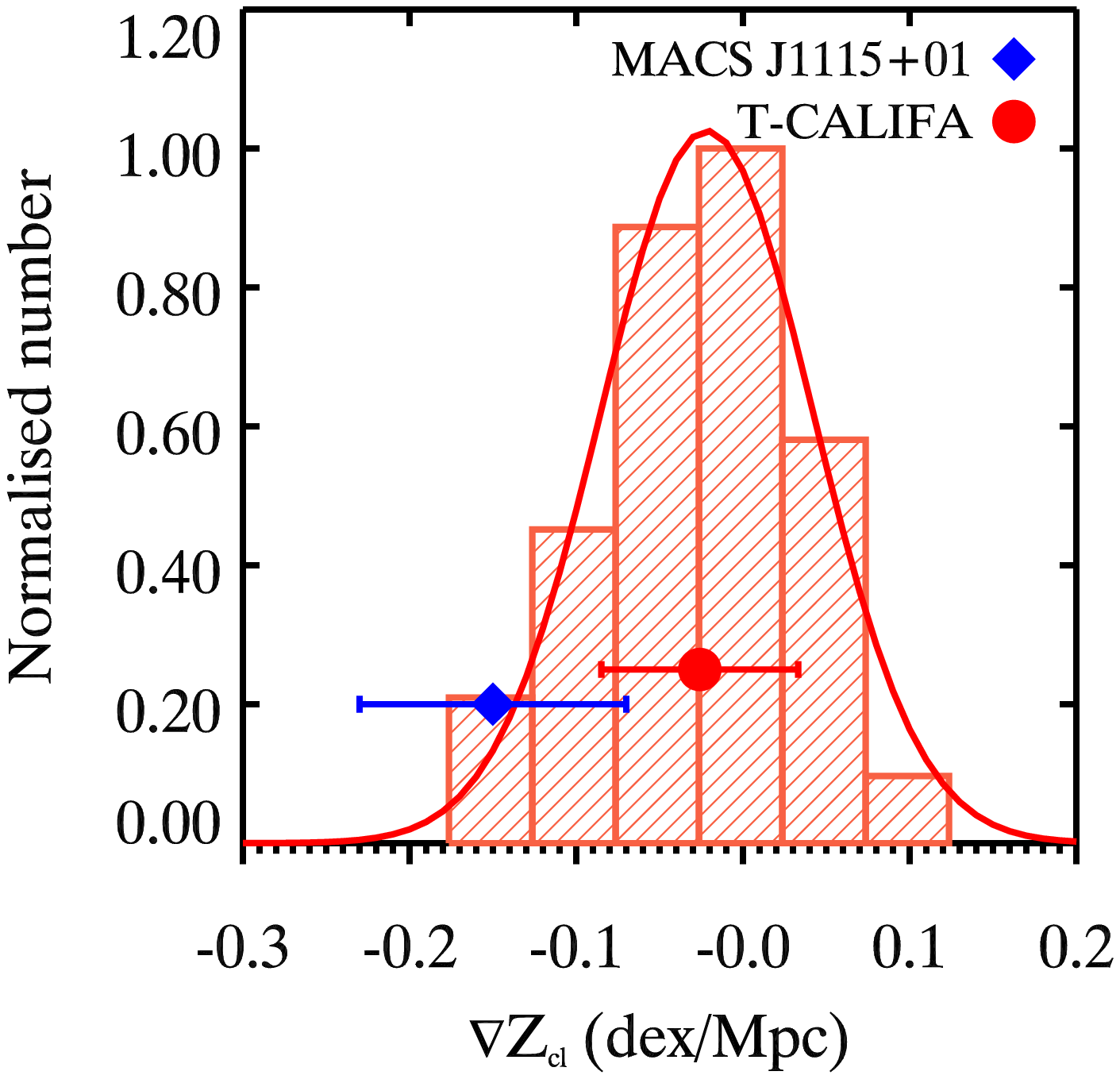}
\caption{ The distribution of the intercept metallicity and the \zcl\  from the disk truncation simulation using full realisations of the  \tcalifa\ sample.  {\bf Left panel:} The intercept distribution with a mean metallicity of $8.69\,\pm\,0.09$ shown as red circle. {\bf Right panel:} The red circle represents the mean (\zcl\,$= -0.03$\,dex/Mpc) and the 1-sigma scatter (0.06\,dex/Mpc) in the \zcl\ distribution  from the disk truncation simulation. The Blue diamond represents \zcl\ observed for MACS\,J1115+01 cluster in Gupta16 (Figure \ref{fig:met_dis}: left panel). The \zcl\ observed for MACS\,J1115+01 is consistent with the \zcl\ produced from disk truncation simulations within 1-sigma errors.}
\label{fig:met_grad_dis}
\end{figure*}

\subsection{Offset in the stellar mass-gas metallicity relation}

The metallicity enhancement produced by disk truncation can be quantified by estimating the offset between the mass-metallicity (MZ) relation of the \tcalifa\ sample and  control samples.  We compare the MZ relation for the \tcalifa\ sample with two control samples:  SDSS7 and \califa\ (\S\ \ref{sec:comp_sample}). We use the logarithmic form proposed by \cite{Zahid2013} parameterised as  
\begin{equation}
12+\log({\rm O/H}) = Z_0-\log\left[1+\left(\frac{M_*}{M_0}\right)^{-\gamma}\right],
\end{equation}
 to fit the MZ relation, where $Z_0$ is the asymptotic metallicity, $M_0$ is the characteristic mass and $\gamma$ represents the power law slope.  The control samples are fit keeping all three parameters free. To estimate the metallicity offset between the \tcalifa\ and control samples, we fit the \tcalifa\ sample as a single parameter model keeping the $M_0$ and $\gamma$ fixed to the comparison control sample's best-fit values. The difference in the asymptotic metallicity ($Z_0$) for the \tcalifa\ sample and the comparison control sample is equivalent to the metallicity offset between two samples. 
 
In Gupta16, we observe a $0.20\,\pm\,0.13$\,dex metallicity enhancement  for cluster members in MACS J1115+01 at a fixed stellar mass (Figure \ref{fig:mz}: left panel).  Figure \ref{fig:mz} shows one realisation of the MZ relation of the \tcalifa\ sample in comparison with the SDSS7 (Figure \ref{fig:mz}: middle panel) and the \califa\ sample (Figure \ref{fig:mz}: right panel). The fit parameters for the MZ relation of the \tcalifa\ galaxies and the referenced control sample are listed in the legend of Figure \ref{fig:mz}.  For a statistical estimate of the metallicity offset between the \tcalifa\ and control samples, we  measure the metallicity offset for each realisation of disk truncation simulation.

Figure \ref{fig:mz_hist} gives the distribution of the metallicity offset between the  \tcalifa\ and the control samples.  Our model predicts a mean metallicity enhancement of $0.02$\,dex for the disk truncated CALIFA galaxies with a 1-sigma scatter of 0.02\,dex on the MZ relation. Disk truncation produces the slightly higher mean metallicity enhancement  between the \tcalifa\ and the SDSS7 sample but within the statistical uncertainties. The truncation of outer-galactic disk due to the RPS produces a 0.02\,dex enhancement in the integrated metallicity of cluster galaxies at a fixed stellar mass. The mean metallicity enhancement from disk truncation is significantly less than the metallicity enhancement observed for MACS J1115+01 in Gupta16 ($0.20\,\pm\,0.13$\,dex).  

\subsection{Cluster-scale metallicity gradient}

In Gupta16, we observe \zcl\,$= -0.15\,\pm\,0.08$\,dex/Mpc in the integrated metallicity of star-forming galaxies for MACS\,J1115+01 (Figure \ref{fig:met_dis}: Left panel). Our disk truncation simulation aims at understanding the contribution of RPS in the \zcl\ observed for MACS\,J1115+01.   We have simulated disk truncation using the IFU datacubes from the CALIFA survey and manually truncate them using a semi-analytic model of RPS.  The right panel in Figure \ref{fig:met_dis} shows the distribution of the integrated metallicity of the \tcalifa\  sample with the cluster-centric distance for one realisation of the disk truncation simulation. We perform robust linear regression to derive the \zcl\ for each realisation of the disk truncation simulation. 

Figure \ref{fig:met_grad_dis} shows the distribution of the intercept metallicity and \zcl\ produced from multiple realisations of the disk truncation simulation (\S\ \ref{sec:sim}). Our disk truncation simulations produce a mean \zcl\ of  $-0.03$\,dex/Mpc with a 1-sigma scatter of $0.06\,$dex/Mpc  in the integrated ISM metallicity of star-forming galaxies. The mean cluster-scale metallicity gradient from the disk truncation simulation  is shallower than the  \zcl\ observed for MACS\,J1115+01, although both are consistent with each other  within 1-sigma standard deviations.  We require measurement of \zcl\ in more galaxy clusters to distinguish the existing \zcl\ distribution from the \zcl\ distribution produced by the truncation of outer-galactic disk. We investigate the origin of the scatter in our simulated cluster-scale metallicity gradient in the following subsection (\S\ \ref{sec:mass_grad}).

\begin{figure}
\centering
\tiny
\includegraphics[scale=0.6, trim=0.5cm 0.5cm 2.0cm 2.0cm,clip=true]{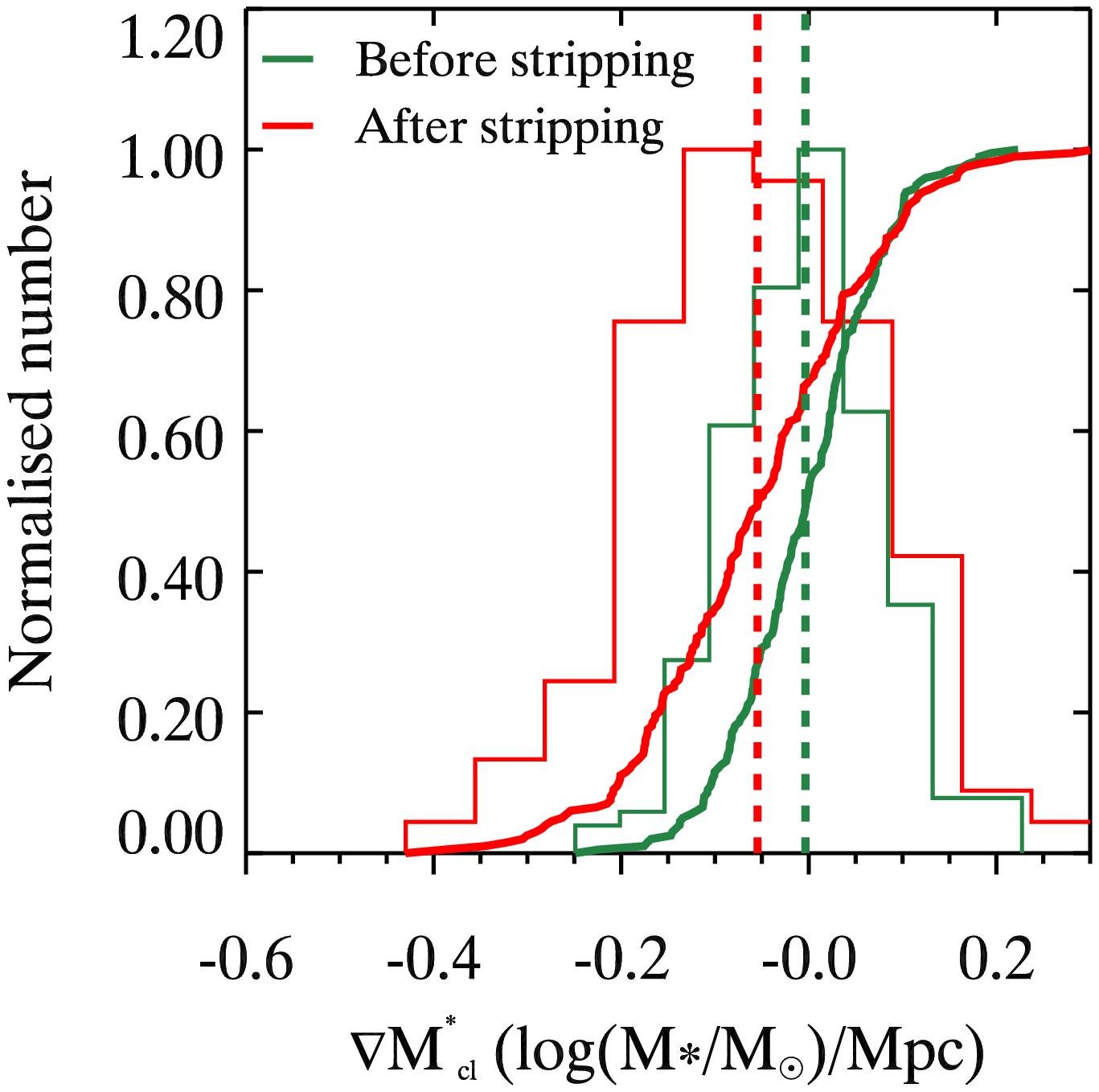}
\caption{The distribution of stellar mass gradient in our  disk truncation simulations. The green and red histograms represent the \mcl\  measured before and after applying the RPS model respectively. The red and green curve represents the corresponding cumulative distribution of the \mcl. The dashed lines show the mean of the two distributions. The disk truncation shifts the mean of the \mcl\ distribution from $0.00\,\log(M_*/M_{\odot})$/Mpc before stripping to $-0.05\,\log(M_*/M_{\odot})$/Mpc after stripping.}
\label{fig:mass_grad_hist}
\end{figure}

\begin{figure*}
\centering
\tiny
\includegraphics[scale=0.5,  trim=0.5cm 0.5cm 0.0cm 1.0cm, clip=true]{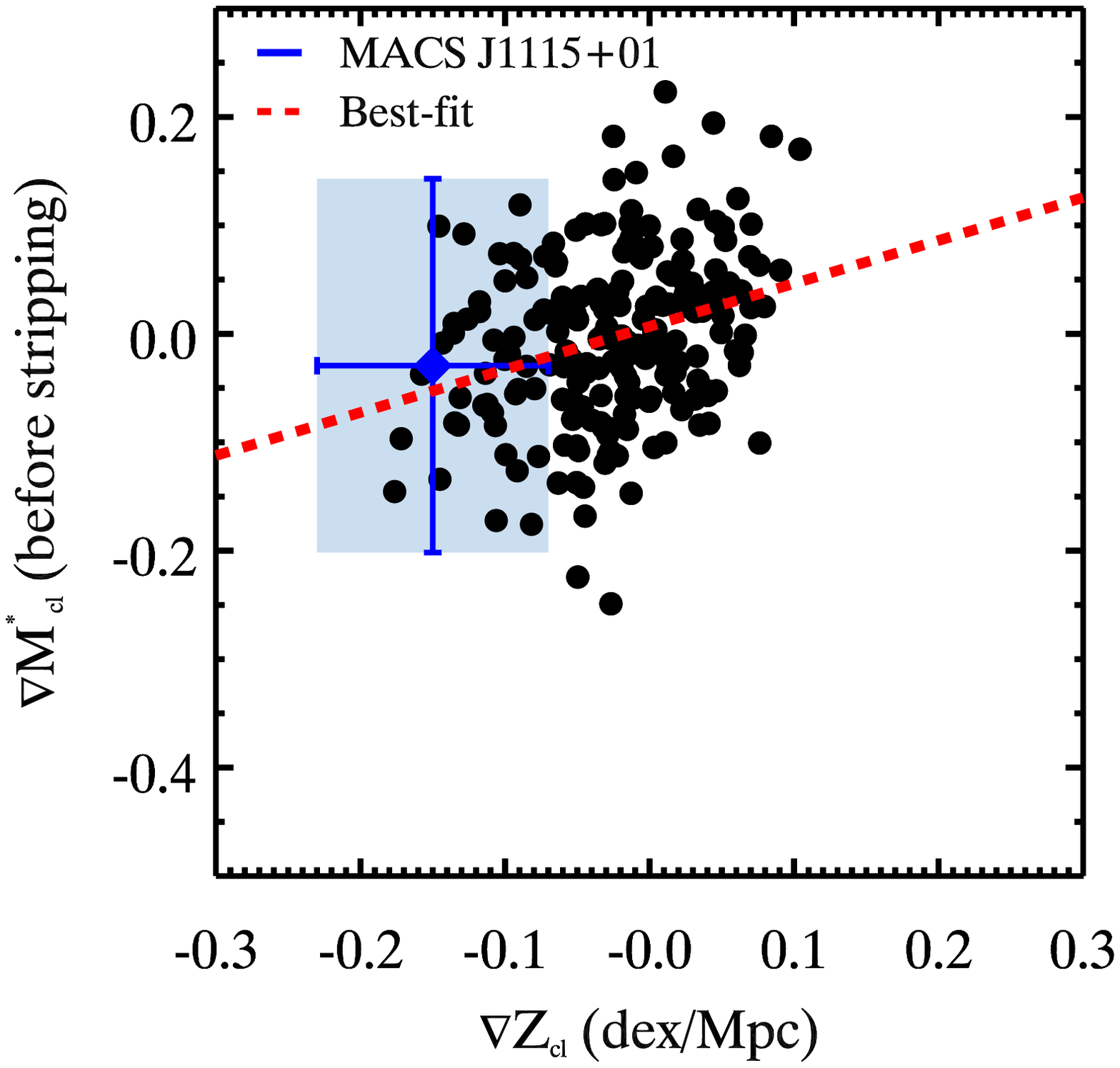}
\includegraphics[scale=0.5,  trim=0.5cm 0.5cm 0.0cm 1.0cm, clip=true]{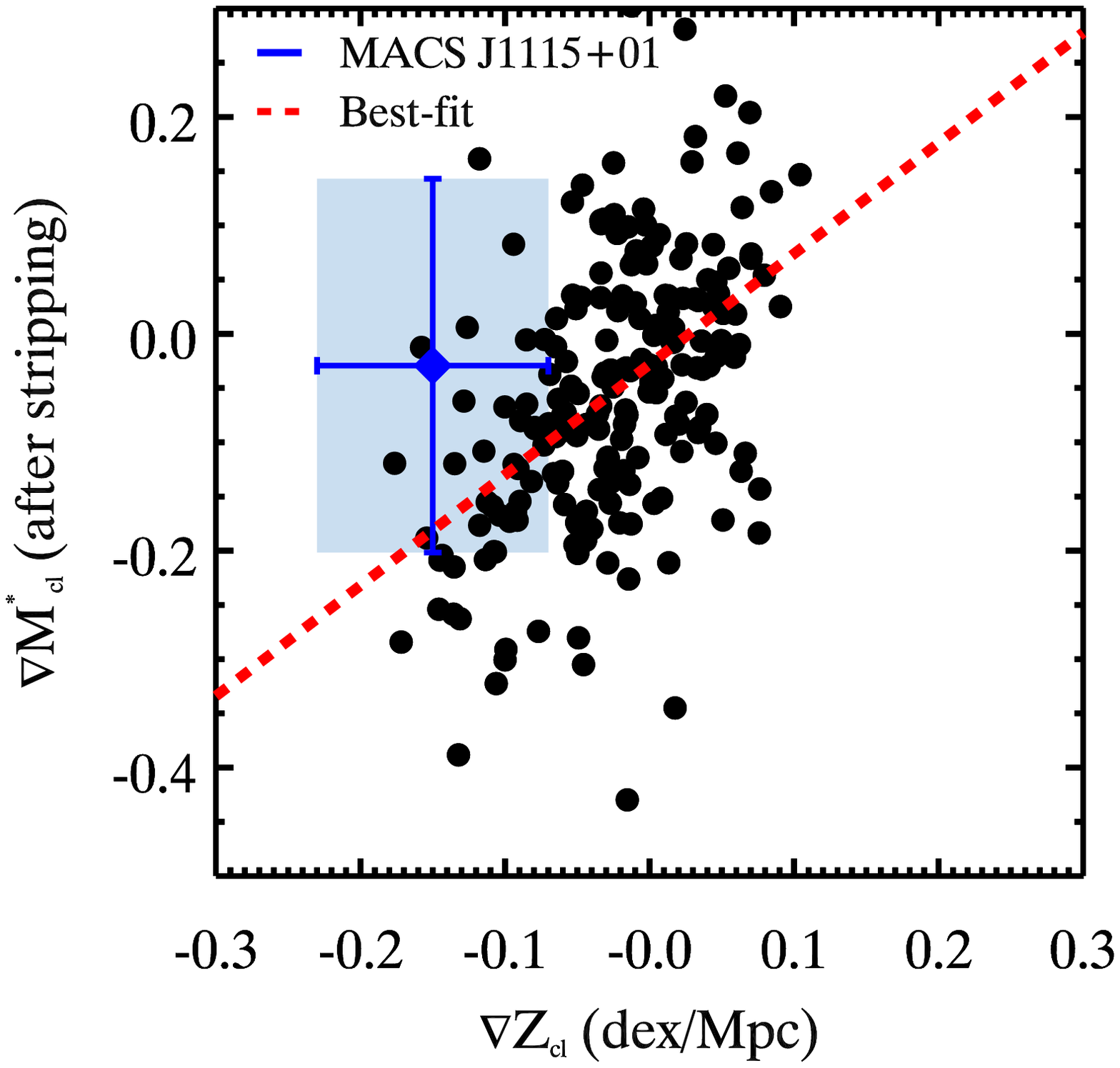}
\caption{The comparison of stellar mass gradient (\mcl) and the estimated cluster-scale metallicity gradient (\zcl). {\bf Left panel:} The \mcl\ measured before stripping versus \zcl. The  \mcl\ in y-axis is the best-fit slope between the stellar mass and cluster-centric distance (see the Appendix). The red dashed line represents the best linear fit and has a slope of $0.39\,\pm\,0.09\,\log(M_*/M_{\odot})$/dex. {\bf Right panel:} The \mcl\ after stripping versus \zcl. The \mcl\ for each realisation is estimated by only including the simulated satellite galaxies with metallicity measurements (see the Appendix). The dashed line represents the best linear fit and has a slope of $1.00\,\pm\,0.12\,\log(M_*/M_{\odot})$/dex. The blue box in each panel represents the mean and 1-sigma measurement error in the \zcl\ and the \mcl\ observed for MACS\,J1115+01 in Gupta16. The \zcl\ from disk truncation simulation is most sensitive to  the \mcl\ introduced by RPS.}
\label{fig:mass_grad}
\end{figure*}

\subsection{The scatter in cluster-scale metallicity gradient}\label{sec:mass_grad}

The predicted cluster-scale metallicity gradient from our disk truncation simulation ranges from $-0.21\,{\rm to\,}+0.18\,{\rm dex/Mpc}$ (Figure \ref{fig:met_grad_dis}: right panel). To simulate cluster galaxies, we uniformly  distribute the mass selected CALIFA galaxies with respect to the cluster-centric distance (\S\ \ref{sec:t-califa}). We define \mcl\  as the best-fit slope between the stellar mass of simulated cluster galaxies and the cluster-centric distance (see the Appendix). Before truncating the galactic disk, we measure a mean \mcl\,$=-0.00\,\log(M_*/M_{\odot})$/Mpc (Figure \ref{fig:mass_grad_hist}) because of the  uniform distribution of selected CALIFA galaxies with respect to the cluster-centric distance. The redistribution of galaxies in the cluster for each realisation introduces a scatter of $0.08\,\log(M_*/M_{\odot})$/Mpc in the \mcl\ before stripping the galactic disks (Figure \ref{fig:mass_grad_hist}). 

The RPS shifts the \mcl\ distribution towards negative values and increases the scatter. The mean \mcl\ after disk truncation is $-0.05\,\log(M_*/M_{\odot})$/Mpc with a 1-sigma scatter of $0.12\,\log(M_*/M_{\odot})$/Mpc (Figure \ref{fig:mass_grad_hist}). The  \mcl\ after truncation is estimated by only including galaxies with measurable metallicities (see the Appendix).  The KS-test yields a probability of nearly zero for the \mcl\ distribution before and after truncation to be derived from the same  distribution. The negative \mcl\ implies that the star-forming galaxies near the cluster center are more massive than the star-forming galaxies in the cluster outskirts. 

Our RPS model does not directly affect the stellar mass of cluster galaxies. The cold gas supply of the low-mass galaxies near the cluster center gets completely stripped of due to RPS.  The star formation in fully gas stripped low mass galaxies is suppressed instantaneously resulting in the complete loss of emission line flux. Therefore, we cannot measure the metallicity of the fully gas stripped low mass galaxies near the cluster center via optical emission lines. A disk truncation simulation realisation with an initial \mcl\ of zero can have a negative \mcl\ after truncation because of the selective removal of the low mass galaxies from the cluster core. 

We find that the \zcl\ predicted by our disk truncation simulation has a direct correlation with the \mcl\ measured after performing disk truncation.  The  \zcl\ shows a weak correlation with the \mcl\ prior to disk truncation (Figure \ref{fig:mass_grad}: left panel). We find a slope of $0.39\,\pm\,0.09\,\log(M_*/M_{\odot})$/dex between the \mcl\ measured before disk truncation and the cluster-scale metallicity gradient. On comparing the \zcl\ with the \mcl\ measured after disk truncation, we find a significantly  steeper slope of $1.00\,\pm\,0.13\,\log(M_*/M_{\odot})$/dex (Figure \ref{fig:mass_grad}: right panel). The best-fit slope of $1.00\,\log(M_*/M_{\odot})$/dex  between the   \zcl\ and the \mcl\ after disk truncation imply that the cluster-scale metallicity gradient in our disk truncation simulation is most sensitive to the \mcl\ introduced by RPS. The negative \mcl\ introduced by the RPS results in a negative \zcl\ because massive galaxies are also metal-rich. 

The \mcl\ observed in Gupta16 for MACS\,J1115+01 is $-0.03\,\pm\,0.17\,\log(M_*/M_{\odot})$/Mpc.  The large measurement error in the \mcl\ for MACS\,J1115+01 does not rule out that the observed \zcl\ is induced by \mcl.  Almost 85\%  of  disk truncation realisations lie outside the 1-sigma range of the \zcl\ and the \mcl\ observed for MACS\,J1115+01 in Gupta16 (Figure \ref{fig:mass_grad}: right panel). To separate the effect of \mcl\ from the effect of \zgal,  we select disk truncation realisations with \mcl\,$=0.00$ after truncation.  We estimate a mean  \zcl\ of $-0.01$\, dex/Mpc with 1-sigma scatter of $0.05$\,dex/Mpc for disk truncation realisations with zero \mcl\ after truncation.  The stellar mass independent \zcl\ is statistically consistent with the \zcl\,$=-0.05\,$dex/Mpc predicted in Section \ref{sec:fiducial} for a cluster galaxy with \zgal\   equivalent to the characteristic \zgal\ of CALIFA galaxies ($-0.1\,{\rm dex/}r_e$; Figure \ref{fig:fid_met_grad}).  The negative cluster-scale metallicity gradient in our disk truncation simulation  is caused by the negative gradient in stellar mass induced by the selective removal of low mass galaxies from the cluster core.

\section{Discussion}\label{sec:discussion}

\subsection{Assumptions in disk truncation model and theoretical uncertainties}\label{sec:assumptions}
Here, we discuss some of the inherent assumptions in our RP model and their potential effect on the RP strength and thus the metallicity gradient measurements:

\begin{itemize}

\item {\bf Effect of inclination angle:} The basic equation of RP stripping by \cite{Gunn1972} assumes that galaxies are falling face-on in the cluster, i.e., the infalling galaxies experience maximum RP. Hydrodynamical simulations of RP stripping suggest that the inclination angle does not play a significant role as long as the galaxies are not moving at an almost edge-on geometry \citep{Roediger2006, Jachym2009, Steinhauser2016}. The total mass loss due to stripping is not affected by the inclination angle but the mass-loss rate is highest for galaxies falling face-on in the cluster \citep{Steinhauser2016}. Hence, we are safe to neglect the role of inclination angle on RP stripping strength. 

\item {\bf Sharp cut-off in radial gas profile:}  Our model assumes that all the cold gas outside the $r_{\rm strip}$ will be removed instantaneously, causing a sharp cut-off in the radial profile of cold gas. Most semi-analytic models with a sharp cut-off in cold gas profile overproduce the fraction of quenched galaxies in the cluster \citep{Guo2013}, suggesting an overestimation of the RP stripping effect. Recent simulations by \cite{Luo2016} relaxes this assumption by assuming that only a fraction of cold gas is removed outside of the stripping radius. However, the continuous cold gas depletion models also fail to improve the over-estimation of the quenched galaxy fraction. Our model uses a sharp cut-off in the radial cold gas profile, thus our results correspond to maximum RP stripping.

\item {\bf Galactic orbits in the cluster:}  The \cite{Gunn1972} formalism assumes that all galaxies are falling into the cluster in radial orbits and are on their first pass through the cluster.  Three-dimensional hydrodynamical simulations suggest that the RPS strength depends on the orbit  of the infalling galaxies. \cite{McCarthy2008} show that galaxies with peripheral orbits undergo a single RP stripping event whereas galaxies with radial orbits undergo multiple RP stripping events depending on the number of pericentric passage. However, more than 80\% of the total mass-loss occurs on the first passage. \cite{Steinhauser2016} show that galaxies with extreme radial orbits undergo significant RPS, whereas galaxies in peripheral orbits only experience mild stripping of gas mass. Thus, the assumption about the galactic orbit would significantly effect the average RPS experienced by satellite cluster galaxies. Our RP prescription assumes that all satellite galaxies have radial orbits, thus providing a upper bound to RP stripping in clusters.

\item {\bf Instantaneous gas stripping:} The RP model assumes instantaneous stripping of cold gas and that the surface gas profile inside the stripping radius remains unaltered.  The metallicity measurements via optical emission lines are sensitive only towards the star-forming regions. The RPS timescales are of the order of $t_{\rm strip} \sim 10-100$\,Myr, whereas the depletion timescale of cluster galaxies are $t_{\rm dep}\sim 1$\,Gyr. Thus, the star-formation is instantaneously suppressed  in the truncated region of the galaxy because $t_{\rm strip} << t_{\rm dep}$. The star-formation inside the stripping radius continues at least till $t\sim t_{\rm dep}$. Relaxing the instantaneous stripping assumption would not significantly affect the RP strength and its effect on integrated metallicity.

\item {\bf ICM density profile:} To describe the ICM density, we use the re-scaled NFW density profile, re-scaled based on the baryonic fraction and the kinetic energy \citep{Makino1998}. \cite{Tecce2010} show that the analytic form for the ICM density overestimates the RP particularly for galaxy cluster at $z\ge 0.5$. The analytic form of the ICM density assumes that the galaxy cluster are relaxed dynamical structures that might not be true particularly at high redshift. Our simulated galaxy cluster is at $z=0.35$ and is a dynamically relaxed system as suggested by the X-ray observations \citep{Allen2007}. The analytic form is expected to provide a good description of the density profile of the cluster. The stripping radius changes logarithmically with the ICM density $r_{\rm strip} \propto \ln(\rho_{\rm ICM})$, i.e.,  \rstrip\ depends weakly on the ICM density. Changing the ICM density profile would not significantly affect the cluster-scale metallicity gradient distribution.

\item {\bf Stellar and gas disk scale length:} To model the surface mass density of the stellar and gaseous component of galaxies, we assume that both the stellar disk and the gaseous disk have the same scale length.  Neutral HI gas disks can extend up to twice the stellar disk size \citep[][and references therein]{Swaters2005, Bosma2016}, but the cold gas disk closely follows  the stellar disk  \citep{Bigiel2012, Cormier2016}. However, the stripping radius in Equation \ref{eq:r_strip} is directly proportional to the scale length of the cold gas disk, therefore uncertainty in the gas disk scale length can significantly affect the stripping radii. We assume same scale length for both gas and stellar disk because otherwise the RPS model would not have an analytic solution.

\item {\bf Radial distribution of cluster galaxies:} To setup our disk truncation simulation, we assume that galaxies are uniformly distributed across the cluster radius. The surface density distribution of both red and blue satellite galaxies follows the NFW profile, albeit with different concentration parameters \citep{Collister2005, Budzynski2012, Zenteno2016}. The RPS can modify the radial distribution of star-forming galaxies in the cluster. Ascertaining the radial distribution of star-forming galaxies in clusters without RP is non-trivial both observationally and via simulations. A uniform radial distribution provides a first order initial condition for the RPS to act upon. Choosing a different initial radial density distribution without invoking the stellar mass segregation would not significantly affect the predicted cluster-scale metallicity gradient distribution. 

 A uniform  radial distribution underestimates the average RPS particularly at radii close to the cluster center. Underestimating the average RPS can potentially affect the mean metallicity enhancement of the MZ relation. To assess the effect of the  radial distribution on the MZ relation, we weight galaxies according to the radial distance while fitting the MZ relation. Galaxies near the cluster center have higher  weights than galaxies in the outskirts. We do not observe a significant change in the mean and 1-sigma scatter of the metallicity enhancement. Thus, the effect of RPS on the gas-phase metallicity is unaffected by the radial distribution of galaxies in the cluster.

\item {\bf Redshift evolution of internal properties of galaxies:} We neglect any redshift evolution in the stellar mass function of field galaxies and \zgal\ by matching the stellar mass function with the  SDSS DR7 data at $z\sim 0$ and simply redshifting the CALIFA galaxies to $z_{\rm cl}$.  The shape of the stellar mass function does not change significantly at intermediate redshift but the characteristic galaxy mass is higher in the high redshift clusters than in the local galaxy clusters or field galaxies \citep{Vulcani2011}. An absolute shift in the characteristic stellar mass would not affect the cluster-scale metallicity gradient.   Observations of local field galaxies show that all isolated spiral galaxies have the same internal metallicity gradient within $\pm 0.14\,{dex/}R_{25}$  \citep{Ho2015, Sanchez2016}. However, the internal metallicity gradient of spiral galaxies evolves with redshift \citep{Yuan2013, Jones2013a, Wuyts2016}. In our simulation, we redshift the CALIFA galaxies from an average redshift of $z=0.01$ to $z_{\rm cl} = 0.35$. This assumes no metallicity gradient evolution till redshift of $z_{\rm cl} = 0.35$, which might not be true. If the internal metallicity gradient of infalling galaxies steepens with redshift then the  cluster-scale metallicity gradient distribution from disk truncation would shift towards more negative values (see \S\ \ref{sec:fiducial}).

\end{itemize}

Considering all assumptions incorporated in our disk-truncation model and the results from recent hydrodynamical simulations, the analytic model of RP stripping used in semi-analytic models corresponds to maximum stripping strength \citep{Steinhauser2016, Natarajan2017}. The cluster-scale metallicity gradient estimated via our disk-truncation model provides an upper-bound to cluster-scale metallicity gradient produced purely by RP stripping. 

\subsection{Origin of cluster-scale metallicity gradient in the RPS simulation}

RPS introduces a mean negative gradient in the stellar mass  of galaxies with the cluster-centric distance (Figure \ref{fig:mass_grad_hist}). RP completely strips off the cold gas supply of the low stellar mass galaxies in the cluster core, i.e., only high stellar mass galaxies continue to form stars in the cluster core.   Numerical simulations of RPS also conclude that the gas fraction lost due to stripping is higher for low mass galaxies \citep{Bekki2009}. The metallicity measurements via optical emission lines are sensitive only to  star-forming galaxies.  Thus, selective quenching of the low mass galaxies in the cluster core introduces a  negative gradient in the stellar mass of star-forming galaxies. The mean negative cluster-scale metallicity gradient in our disk truncation simulation is caused by the negative stellar mass gradient (Figure \ref{fig:mass_grad}: right panel).

Our simulation predict a  nearly flat  \zcl\ purely by the stripping of outer galactic disk that is inconsistent with the \zcl\ in observations ($-0.01$\,dex/Mpc compared to $-0.15\,\pm\,0.08$\,dex/Mpc).  Our maximal RPS model predicts  significantly smaller metallicity enhancement of the MZ relation than observations ($+0.02$\,dex compared to $+0.20$\,dex).  We suspect that additional physical mechanisms such as self-enrichment via strangulation might be necessary to fully reproduce the observed cluster-scale metallicity gradient and metallicity enhancement of the MZ relation. Observations of more galaxy clusters are required to fully sample the existing cluster-scale metallicity gradient distribution  and differentiate  from the metallicity gradient distribution produced purely by disk truncation.

\section{Conclusion}

The truncation of the outer-galactic disk in the cluster environment due to RPS has been used to explain the enhanced metallicity of cluster galaxies compared to  counterpart field galaxies \citep{T.M.Hughes2012}.  In Gupta16, we present the first observation of a cluster-scale gradient in the ISM metallicity of star-forming cluster members. We hypothesize that the self-enrichment of cluster members via strangulation and/or the truncation of outer-galactic disk can produce a negative gradient  in the integrated metallicity of  cluster members. In this paper, we simulate disk-truncation using  an analytic model of RPS. We use a stellar-mass matched sample of CALIFA galaxy (Figure\,\ref{fig:mass_dis}) to create a mock sample of disk truncated cluster galaxies. We compare the prediction from our disk truncation simulations with the observations from Gupta16 by conducting mock DEIMOS observations.   

Our model of disk truncation results in a  mean cluster-scale metallicity gradient of $-0.03$\,dex/Mpc with a 1-sigma scatter of $0.06\,$dex/Mpc (Figure\,\ref{fig:met_grad_dis}: right panel).  The cluster-scale metallicity gradient observed for MACS\,J1115+01 in Gupta16 lies within the cluster-scale metallicity gradient distribution produced by our disk truncation simulation (Figure\,\ref{fig:met_grad_dis}: right panel).  We estimate a metallicity enhancement of   $+0.02$\,dex  by comparing the MZ relation of the selected CALIFA galaxies before and after disk truncation (Figure \ref{fig:mz_hist}: right panel). The metallicity enhancement from disk truncation is significantly smaller than the metallicity enhancement observed for MACS\,J1115+01 ($+0.20$\,dex). The random distribution of galaxies in the clusters produces a $0.08\,\log(M_*/M_{\odot})$/Mpc scatter in the stellar mass gradient of cluster members. The scatter in stellar mass distribution is the primary source of scatter in the cluster-scale metallicity gradient estimated in our disk truncation simulation. 

RPS induces a negative gradient in the stellar mass of star-forming galaxies (Figure \ref{fig:mass_grad_hist}). The stellar mass of a galaxy remains unaffected in our RPS model, but RPS leads to complete loss of the gas supply of low mass galaxies in the cluster core. Only massive galaxies continue to form stars in the cluster cores, resulting in a negative gradient in the stellar mass of star-forming cluster galaxies with cluster-centric distance. The cluster-scale metallicity gradient predicted in our disk truncation simulation is most sensitive to the stellar mass gradient induced by the RPS (Figure \ref{fig:mass_grad}: right panel). The mass segregation induced by the selective removal of low mass galaxies drive the negative cluster-scale metallicity gradient predicted in our simulation.

The stellar mass independent cluster-scale metallicity gradient predicted by our disk truncation simulation is $-0.01\,$dex/Mpc with a 1-sigma scatter of $0.05\,$dex/Mpc, which is significantly smaller than the gradient of $-0.15\,\pm\,0.08\,$dex/Mpc observed in MACS\,J1115+01. Some of the assumptions in our RPS model such as the sharp cut-off in the radial gas profile and single infall in radial orbits only maximize the strength of RP in galaxy clusters (\S\ \ref{sec:assumptions}).  Thus, the truncation of outer galactic disk via RPS alone can not explain the cluster-scale metallicity gradient observed in MACS\,J1115+01.

Observations of cluster-scale metallicity gradients in more galaxy clusters are required to  characterise the full range of cluster-scale metallicity gradients existing in galaxy clusters. A flat mean cluster-scale metallicity gradient and the small metallicity enhancement of the MZ relation in our disk truncation simulation suggest that additional physical mechanisms such as strangulation are likely to be required to reproduce the observed cluster-scale metallicity gradient.  In the future, we plan to incorporate a simple chemical evolution model with stripped neutral Hydrogen to understand the effect of strangulation on the integrated cluster member metallicities.

\section*{Acknowledgements}

The author thank the referee for providing useful comments and suggestions to improve the quality of the paper. This study uses data provided by the Calar Alto Legacy Integral Field Area (CALIFA) survey (\url{http://califa.caha.es/}). Based on observations collected at the Centro Astron\'omico Hispano Alem\'an (CAHA) at Calar Alto, operated jointly by the Max-Planck-Institut f\H{u}r Astronomie and the Instituto de Astrof\'isica de Andaluc\'ia (CSIC). A.G.  acknowledges I-Ting Ho for help with the CALIFA data and Melanie Kaasinen for  her help in the editing of this paper.   L.J.K. gratefully acknowledges support from an Australian Research Council (ARC) Laureate Fellowship (FL150100113). K. Tran acknowledges support by the National Science Foundation under Grant \#1410728. Davide Martizzi was supported by the Swiss National Science Foundation as an Advanced Postdoc.Mobility Fellow until November 2016; grant number P300P2\_161062.

\bibliographystyle{aasjournal}

\begin{appendix}\label{ap:mass_grad}

\begin{figure*}
\centering
\tiny
\includegraphics[scale=0.35, trim=0.0cm 1.0cm 0.0cm 0.0cm,clip=true]{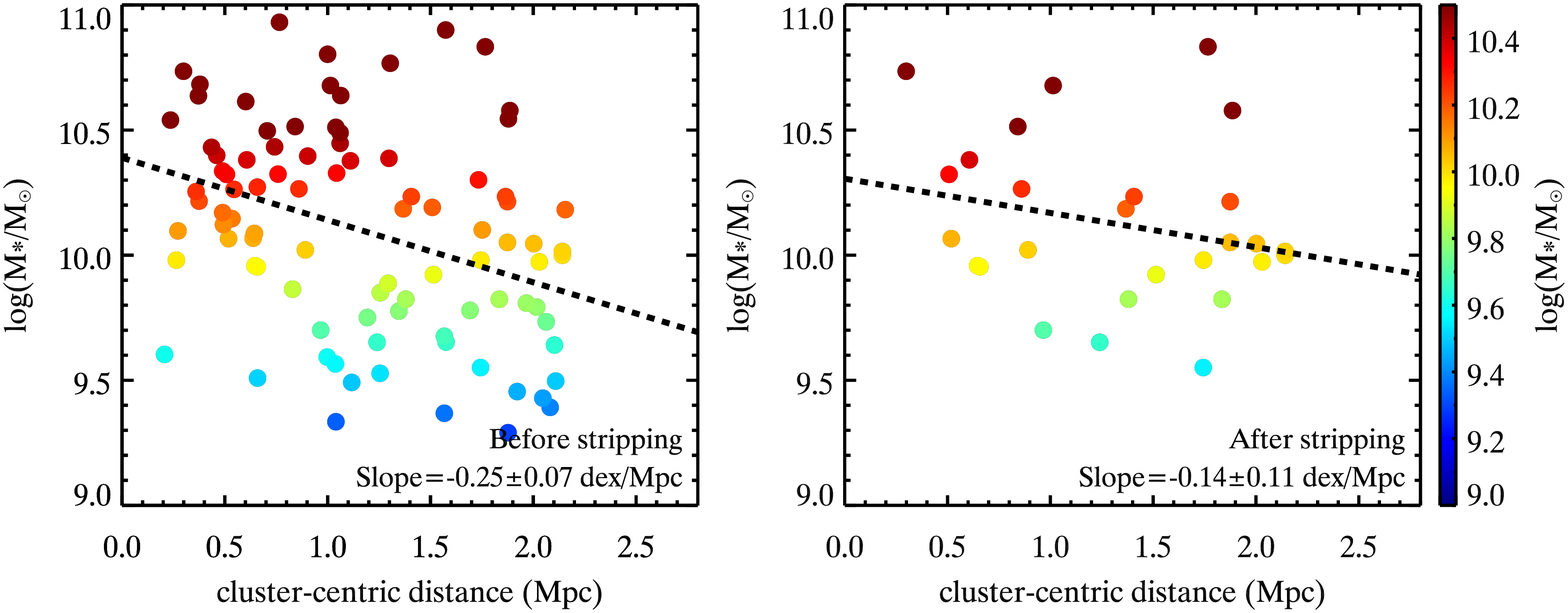}
\includegraphics[scale=0.35, trim=0.0cm 1.0cm 0.0cm 0.0cm,clip=true]{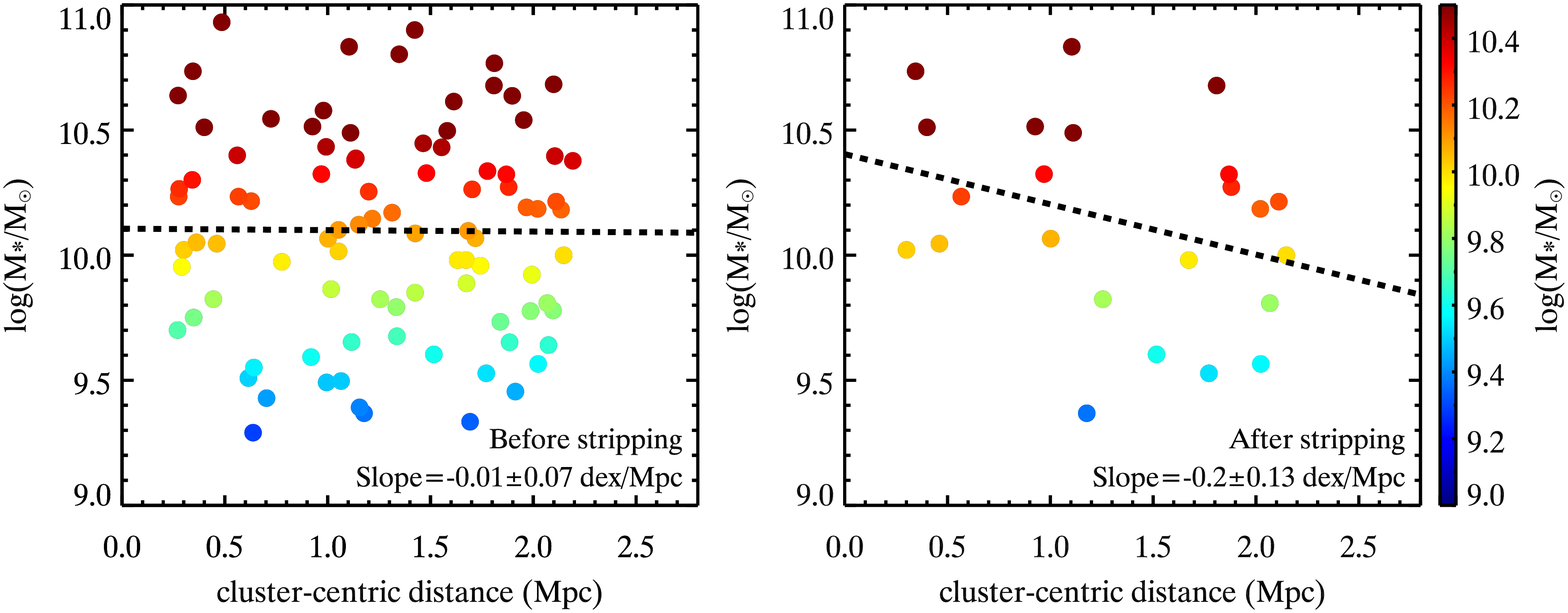}
\includegraphics[scale=0.35, trim=0.0cm 1.0cm 0.0cm 0.0cm,clip=true]{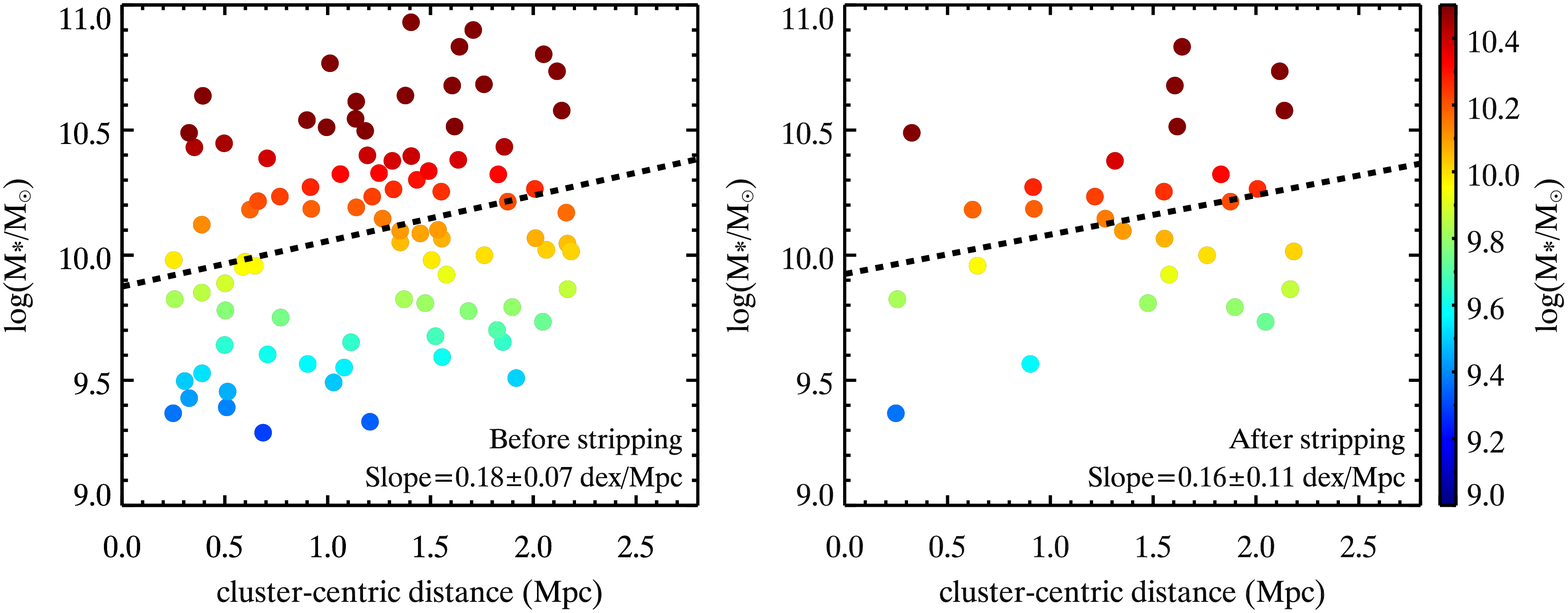}
\caption{Examples demonstrating the estimation of \mcl\ for different realisations of disk truncation simulations. The left  column shows the stellar mass distribution with the cluster-centric distance before implementing the RPS model. The right column shows the  stellar mass distribution of truncated  cluster galaxies with measurable gas-phase metallicities. The dashed line in each panel represents the best linear fit to the data. The data points in each figure are color coded with stellar mass. The \mcl\ before stripping is positive in first row, zero in the middle row and negative for the last row. The \mcl\ for the first and the last row remains statistically the same after stripping but the \mcl\ shifts to negative value for the middle row after stripping.   }
\label{fig:mass_grad_ex}
\end{figure*}

In section \ref{sec:mass_grad}, we compare the stellar mass gradient in our disk truncation simulation with the predicted cluster-scale gradient in integrated metallicity.  The \mcl\ is the best-fit slope between the stellar mass of galaxies and the cluster-centric distance for a single realisation of disk truncation simulation. For each realisation of  disk truncation simulation, we measure a \mcl\ before and after applying the RPS model. The \mcl\ after stripping is measured by only including galaxies for which gas-phase metallicity can be estimated. Figure \ref{fig:mass_grad_ex} shows the stellar mass distribution for three realisations of the disk truncation simulation before and after stripping.   For realisations with an initial \mcl\ equal to either positive or negative, the  \mcl\ after truncation remains statistically unchanged (Figure \ref{fig:mass_grad_ex}: top \& bottom rows). Whereas, the disk truncation realisation with a zero \mcl\  before stripping can have a \mcl\ after stripping (Figure \ref{fig:mass_grad_ex}: middle row). 

\end{appendix}

\end{document}